\documentclass[default, twocolumn, longbib]{aastex701}

\accepted{October 2, 2025}
\submitjournal{AAS Journals}
\begin{document}
\title{Detection of an NH$_3$ absorption band at 2.2 $\mu$m on Europa}
\author[orcid=0000-0002-4420-0595,sname='A.']{A. Emran}
\affiliation{NASA Jet Propulsion Laboratory, California Institute of Technology, Pasadena, CA 91109, USA}
\email[show]{al.emran@jpl.nasa.gov}

\begin{abstract}
The presence of NH$_3$-bearing components on icy planetary bodies has important implications for their geology and potential habitability. Here, I report the detection of a characteristic NH$_3$ absorption feature at 2.20 $\pm$ 0.02 $\mu$m on Europa, identified in an observation from the Galileo Near Infrared Mapping Spectrometer. Spectral modeling and band position indicate that NH$_3$-hydrate and NH$_4$-chloride are the most plausible candidates. Spatial correlation between detected ammonia signatures and Europa’s microchaos, linear, and band geologic units suggests emplacement from the underground or shallow subsurface. I posit that NH$_3$-bearing materials were transported to the surface via effusive cryovolcanism or similar mechanisms during Europa's recent geological past. The presence of ammoniated compounds implies a thinner ice shell \citep{spohn&schubert2003} and a thicker, chemically reduced, high-pH subsurface ocean on Europa \citep{handetal2009}. With the detection of NH$_3$-bearing components, this study presents the first evidence of a nitrogen-bearing species on Europa— an observation of astrobiological significance given nitrogen’s essential role in the chemistry of life.

\end{abstract}

\section{Introduction} 
The detection of ammonia (NH$_3$) or ammonia-bearing\footnote{In this paper, ammonia-bearing (NH$_3$-bearing) refers broadly to any ammoniated species, including ammonia-derived compounds, ammonium-bearing species, or NH$_4$+-bearing materials—that is, attributed to any NH- components in general.} components (NH\textsubscript{3}–hydrate, –salts, or –minerals) on icy planetary bodies in the Solar System is of significant interest for understanding their geology, potential habitability, and astrobiological relevance \citep{kargel1992, bergetal2016, cruikshank2019a}. On Jupiter’s moon Europa, the presence of NH\textsubscript{3} or ammoniated (NH\textsubscript{3}- or NH\textsubscript{4}-bearing) species is particularly important for constraining ocean chemistry, assessing habitability, and reconstructing the moon’s early atmosphere \citep{handetal2009, vanceetal2018, vance2023, moulanieretal2025}. NH\textsubscript{3} acts as an anti-freezer \citep{grasset&sotin1996, neveu2017aqueous}; an abundant presence of this can lower the freezing point of liquid H\textsubscript{2}O by up to 100 K \citep{kargel1992} and may enable retention of subsurface oceans for icy bodies \citep{spohn&schubert2003, trinh2023slow, decolibus2023nh3}. Although it remains unclear whether Europa’s subsurface ocean \citep{khurana1998induced,  pappalardo1999does, kivelson2000galileo} is directly connected to its surface \citep{villanueva2023endogenous} the detection of NH\textsubscript{3}– compounds may suggest such a connection \citep{spohn&schubert2003}, as these materials are unstable in space radiation \citep{strazzulla1998evolution,moore2007ammonia, loeffler2010radiation}. Thus, the presence of NH\textsubscript{3} or ammoniated contents may indicate cryovolcanic activity \citep{neveu2015prerequisites} or other endogenic processes that transported material from the underground ocean or shallow subsurface to the surface in the recent geologic past. Furthermore, detection of an ammoniated content can provide valuable constraints on the composition and chemistry of Europa’s subsurface ocean, which has remained poorly understood to date \citep[e.g.,][]{becker2024exploring}.

NH\textsubscript{3}-bearing components have been confirmed in the icy bodies across the outer solar system through detection of the characteristic absorption band near $\sim$2.20$\mu$m \citep[e.g.,][]{dalle2019detection, cruikshank2019b, emran2023surface, cook2023analysis, bauer2002near, cartwright2020evidence, protopapa2024detection}. Though ammonia has another band at $\sim$2.0$\mu$m, this feature is typically obscured by the strong H\textsubscript{2}O ice absorption in the same spectral region, making reliable identification of ammonia challenging on icy astronomical bodies \citep{zheng2009infrared, roser2021infrared}. Ammoniated components have been confirmed on Pluto and its moons—Charon, Nix, and Hydra \citep{cruikshank2019b, dalle2019detection, brown2000evidence, emran2023surface, cook2007near, cook2023analysis, protopapa2024detection}. On Pluto and Charon, these species are thought to originate from cryovolcanic activity in the recent geologic time \citep{cook2007near, cruikshank2019b, dalle2019detection, emran2025kiladze}. An NH\textsubscript{3} absorption band near 2.20$\mu$m has been detected on the surfaces of Uranian moons, including Ariel, Miranda, Umbriel, Oberon, and Titania \citep{decolibus2023nh3, cartwright2018red, cartwright2020evidence, cartwright2023evidence, bauer2002near}. On Ariel, ammoniated materials have been hypothesized to be emplaced as ammonia-rich cryolava, suggesting geologic activity in the recent past \citep{cartwright2020evidence}. In the Saturn system, ammonia has been detected in plume material from Enceladus \citep{waite2009liquid}. Possible detections of ammoniated species have been reported on the surfaces of Enceladus, Dione, and Iapetus \citep{emery2005near, verbiscer2006near, clark2008compositional, clark2012surface}. Observations from the instruments onboard NASA’s Juno spacecraft confirm possible detection of NH\textsubscript{3}-bearing components on Jupiter’s moon Ganymede \citep{molyneux2022ganymede}, interpreted as likely salt residues originating from its deep subsurface ocean brine that was emplaced onto the surface \citep{tosi2024a}.

Although Europa has long been suspected of hosting ammonia or ammoniated contents \citep[e.g.,][]{lewis1971satellites, Kargel1998, spohn&schubert2003, vanceetal2018, tosi2024b}, definitive evidence has remained elusive \citep{vance2023, becker2024exploring, moulanieretal2025}. Existing studies have reported a weak signal near 2.2$\mu$m at the trailing hemisphere using NASA’s Infrared Telescope Facility (IRTF) data obtained in 1980 and 1985– suspected as a possible signature of NH\textsubscript{3}-hydrate in a weak mixture with H\textsubscript{2}O ice \citep{clark1980galilean, brown1988search}. However, subsequent observation in 1986 with improved instrumentation did not detect this absorption feature, casting doubts about the reliability of previous observations or suggesting that the material responsible for the weak absorption might no longer be detectable \citep{brown1988search}. Another subsequent study using telescopic data \citep{calvin1995spectra} to search for the 2.2$\mu$m band also did not find the absorption feature \citep{carlson2009europa}. In this study, I used an observation from the Near Infrared Mapping Spectrometer \citep[NIMS;][]{carlson1992near} onboard the Galileo spacecraft to identify the characteristic NH\textsubscript{3} absorption band at $\sim$2.20$\mu$m on Europa’s surface. I also compared the spectral features observed in the NIMS data to laboratory spectra of a range of ammonia and ammonia-bearing candidates, including amorphous and crystalline NH\textsubscript{3} ice \citep{roser2021infrared, hudson2022ammonia}, NH\textsubscript{3}-hydrate \citep[NH\textsubscript{3}.H\textsubscript{2}O; ][]{brown1988search}, NH\textsubscript{3}-bearing salts \citep{fastelli2020reflectance, fastelli2022reflectance}, and NH\textsubscript{3}-bearing phyllosilicate. Additionally, I applied linear spectral modeling to deconvolve the compositional mixture of ice and non-ice materials within the detected pixels in the Galileo/NIMS observation.

\begin{figure*}[ht!]
\plotone{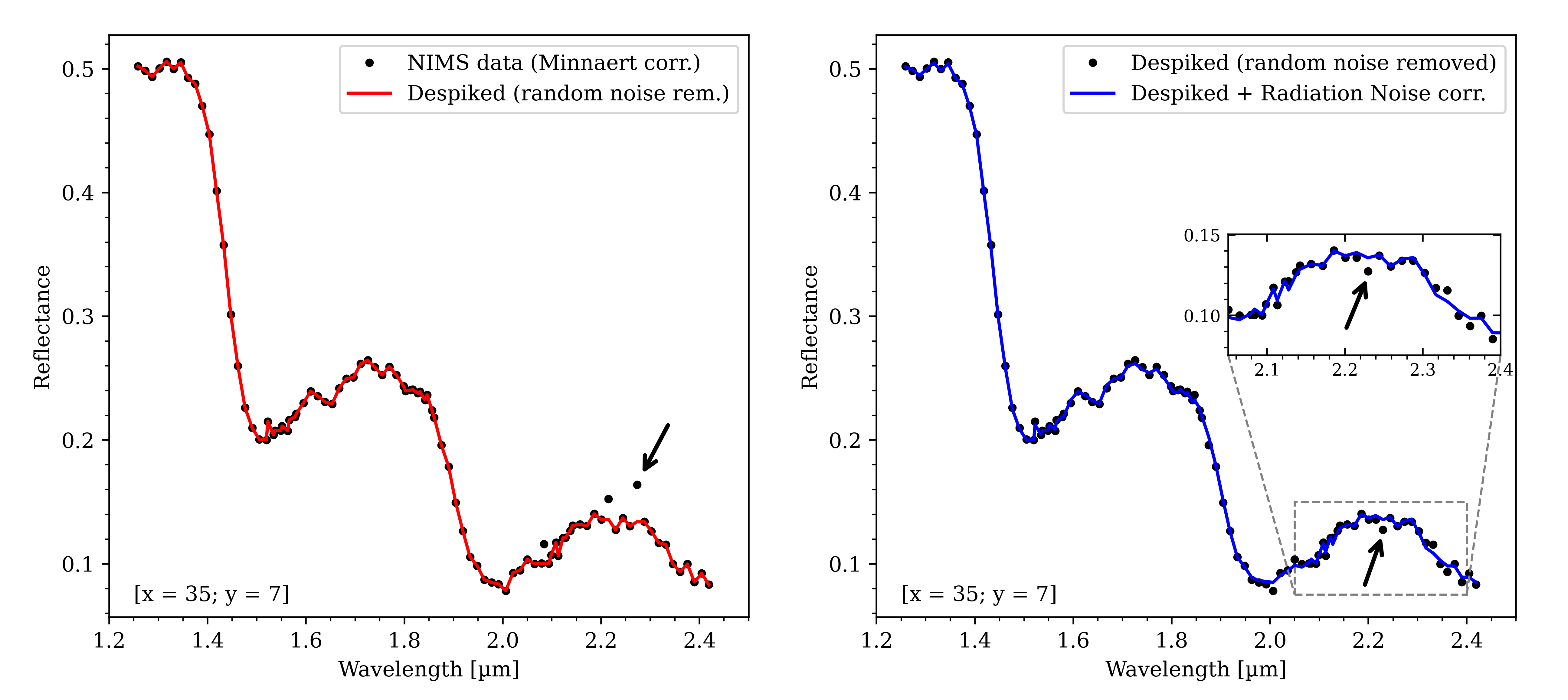}
\caption{An example of noise removal routine applied to the NIMS scan. \textit{Left panel}: Spectral denoising using despiking approach applied to the Minnaert-corrected spectrum (black dots). The black arrow indicates a random noise spike near $\sim$2.3$\mu$m that was removed in the despiked spectrum (red line). Any black dot (Minnaert-corrected data point) not coinciding with the red line (despiked spectrum) is considered a noise artifact and was removed and replaced during the despiking step. \textit{Right panel}: Radiation noise removal via spatial smoothing applied to the despiked spectrum (black dots). The black arrow indicates a single-band noise artifact near $\sim$2.2$\mu$m that was treated in the final smoothed spectrum (blue line). Any black dot (despiked data point) not aligned with the blue line (spatially smoothed spectrum) is considered radiation noise and was corrected during the radiation removal step. Inset is the figure with a zoom on the $\sim$2.05-2.40$\mu$m region. The x, y coordinates are the location of the pixel in NIMS observation 11ENCYCLOD01A of \cite{malaska2024updated}.
\label{fig:fig1}}
\end{figure*}

\section{Observation and Methods} \label{sec:style}
\subsection{Data and calibration} \label{subsec:data}
This study used the Galileo/NIMS observation 11ENCYCLOD01A– consisting of hyperspectral data covering wavelengths from 0.7 to 5.2$\mu$m \citep{carlson1992near}. The image scene covers $\sim$134°11'–148°47'E longitudes and $\sim$11°15'–19°55'N latitudes on Europa’s trailing hemisphere at a spatial (ground) resolution of $\sim$4.7 km/pixel. This observation was selected because it includes microchaos, linear, and band geologic features—among the young geological units on Europa \citep[e.g.,][]{leonard2024global} and represent surface geologic evidence of cryovolcanism or surface-subsurface material exchange \citep{wilson1997eruption, pappalardo1998geological, pappalardo1999does, daubar2024planned}. I analyzed the radiance factor or reflectance product derived from the NIMS data that was recently reprocessed and geospatially co-registered to Europa’s surface by \cite{malaska2024updated}. The raw NIMS experimental data records (EDR) were calibrated and processed using standard routines, including photometric correction and conversion to reflectance units. Refer to \cite{malaska2018europa, malaska2024updated} for details of the Galileo/NIMS processing routine. I utilized the Minnaert-corrected version of the dataset \citep{minnaert1941reciprocity, mcewen1991photometric, verbiscer1998reflectance}, which accounts for variations in observed reflectance due to changing observation geometries \citep{shirley2010europa}.

\begin{figure*}[ht!]
\plotone{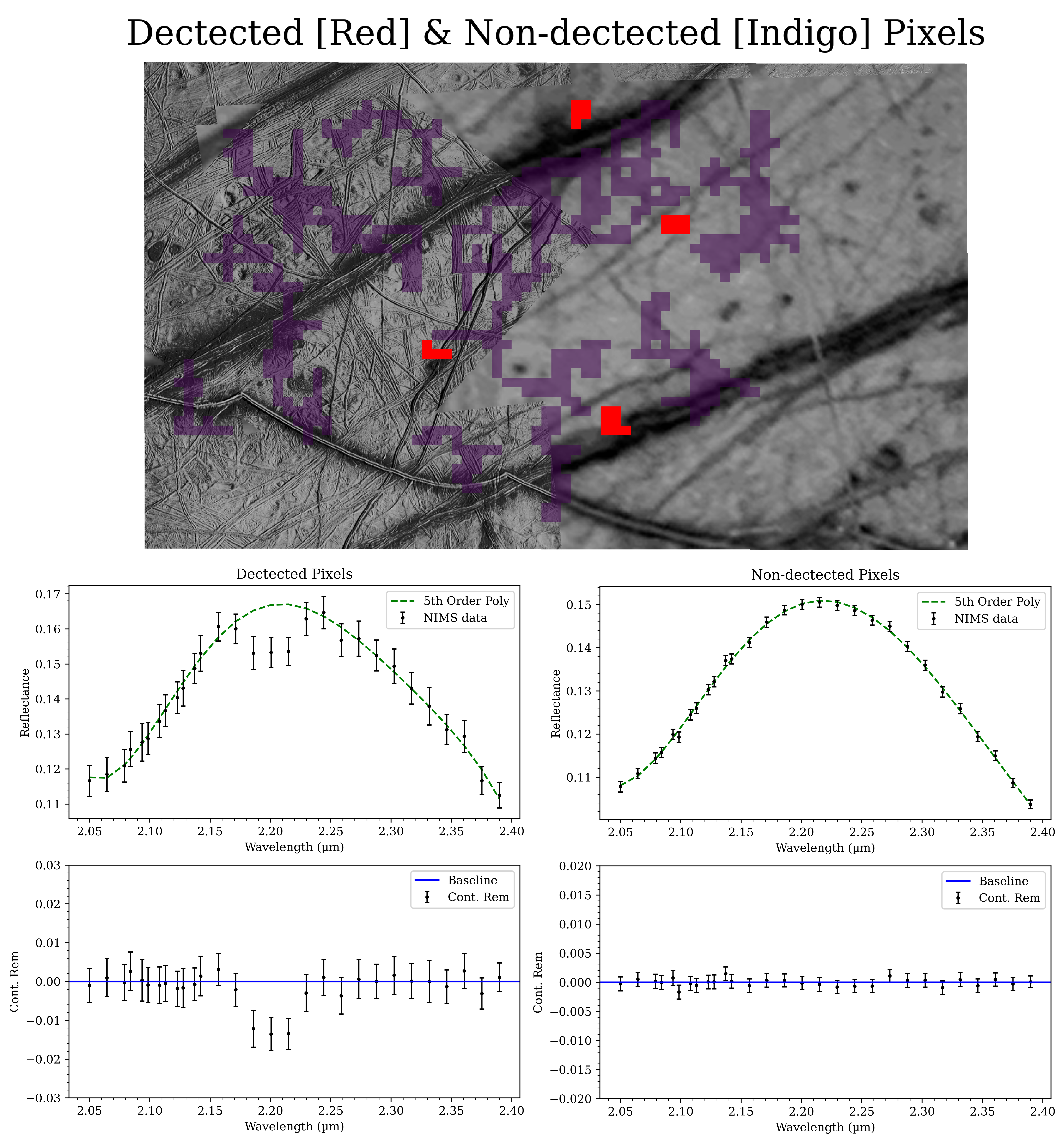}
\caption{\textit{Upper panel}: Locational distribution of detected pixels (red polygons) and some regions of non-detected pixels (indigo polygons) using 5\textsuperscript{th}-degree polynomial continuum fit from Galileo/NIMS observation 11ENCYCLOD01A overlaid on the Galileo/SSI basemap \citep{malaska2018europa, malaska2024updated}. \textit{Middle panel}: The continuum fits to the average reflectance spectra of detected pixels (left subplot) and some non-detected pixels (right subplot) in the NIMS scan. \textit{Lower panel}: The continuum-removed average spectra along with associated standard errors against the continuum baseline for detected pixels (left subplot) and some non-detected pixels (right subplot). For detected pixels, the reflectance and standard errors values at 2.19-2.21$\mu$m fall clearly below the continuum baseline— confirms the feature exceeds the noise level.
\label{fig:fig2}}
\end{figure*}

In this instance, I used a subset of the NIMS data covering the 1.2–2.5$\mu$m range at a spectral resolution of about 0.025$\mu$m \citep{carlson1992near}. Although the reflectance product was fully processed, it still retains residual random and radiation noises \citep{shirley2010europa}. To mitigate this, I applied a two-step noise removal procedure. First, I performed spectral denoising by applying a median filter along the spectral axis, independent of pixel location. Using a filter window of three bands, I identified and replaced any spectral values exceeding 3$\sigma$ with the median of the bin– a treatment used for random noise. This method removes sudden (abrupt) noise spikes from spectra— an approach commonly referred to as despiking. An example result, shown in the left panel of Fig. \ref{fig:fig1}, demonstrates how the despiking process removed a sharp noise artifact around $\sim$2.3$\mu$m (indicated by the arrow) from the Minnaert-corrected spectrum. Note that this step also removes similar abrupt spikes at other wavelengths, as shown in Fig. \ref{fig:fig1}. Accordingly, this routine effectively removes random noise from spectra across the entire NIMS scan— returning much cleaner spectra compared to the original Minnaert-corrected data. 

Second, I applied a 3×3-pixel spatial smoothing (average) filter to each spectral slice, further reducing noise while preserving the underlying signal. This step is particularly effective at mitigating radiation noise. As shown in the right panel of Fig. \ref{fig:fig1}, spatial smoothing routine successfully removed a single-channel spurious feature near $\sim$2.2$\mu$m (indicated by the black arrow), reinforcing the robustness of absorption features that span multiple spectral channels at a given pixel location. Accordingly, this step effectively identifies and corrects spurious single-channel artifacts caused by radiation noise across the NIMS scan. The use of spatial smoothing improved the spectral quality by a factor of three compared to the Minnaert-corrected and despiked NIMS spectra at each pixel– an improvement in radiation noise \citep{daltonetal2012}. Thus, the combined spectral and spatial cleaning routine used in this study has significantly reduced noise uncertainties in the NIMS spectra. An example of these corrections applied to the 2.36$\mu$m reflectance band of the image observation is shown in Fig. \ref{fig:fig8}, and an example of a corrected spectrum at a single pixel location is given in Fig. \ref{fig:fig9}.

\subsection{Identification of 2.2$\mu$m band} \label{subsec:identification}
To detect NH\textsubscript{3}-bearing species, I isolated the $\sim$2.20$\mu$m spectral region from the noise-corrected NIMS data by extracting the spectrum spanning 2.05 - 2.40$\mu$m across the entire NIMS scan. I applied a fifth-degree polynomial continuum fit to each pixel’s spectrum between 2.05 and 2.40$\mu$m, excluding the 2.17–2.23$\mu$m region to preserve the target absorption feature. This approach is similar to methods used to detect organics and CO\textsubscript{2} on icy moons in the Saturn system and Jupiter's Trojans using JWST data \citep{belyakov2025saturnian, wong2024jwst}. Band depth was calculated as the average of relative absorption at 2.19, 2.20, and 2.21$\mu$m channels from the continuum fit at each pixel. To ensure robust detection, I considered only those pixels with a band depth greater than 0.01 (\(>1\%\)).

To eliminate spurious isolated detections, only clusters where pixels are contiguous (directly connected) of at least three pixels were considered. For each cluster, I extracted and visually inspected the average spectrum to confirm the presence of the $\sim$2.20$\mu$m absorption feature and exclude false positives. This series of rigorous procedures ensures reliable detection and minimizes the influence of spectral artifacts. After verifying the true detection clusters, I averaged the reflectance spectrum of all pixels from the verified clusters and estimated the standard errors for the corresponding wavelengths. The standard error of the average reflectance was calculated as: \[\sigma_E = \frac{\sigma}{\sqrt{N}}
;\] where \(\sigma_E\) is the standard error, \(\sigma\) is the standard deviation of the reflectance, and \(N\) is the number of detected pixels \citep{altman2005standard}. I found that the calculated mean standard error is $\sim$3\% over the wavelengths of the average reflectance spectrum of detected pixels—comparable to the uncertainty (±3–5\%) used in previous compositional studies using Galileo/NIMS data \citep{daltonetal2012}.

I compared the continuum fit to the average reflectance spectra of detected pixels and some regions of non-detected pixels in the NIMS scan to verify the use of the 5\textsuperscript{th}-degree polynomial continuum fit for identifying the 2.20$\mu$m absorption band (upper panel of Fig. \ref{fig:fig2}). For detected pixels, the reflectance at 2.19, 2.20, and 2.21$\mu$m falls noticeably below the fitted continuum, which otherwise closely follows the surrounding wavelengths from 2.05 to 2.40$\mu$m (middle panel, Fig. \ref{fig:fig2}). In contrast, for non-detected pixels, the continuum fits closely across the entire spectral range, including the 2.19–2.21$\mu$m region (middle panel, Fig. \ref{fig:fig2}). This comparison validates the application of a 5\textsuperscript{th}-degree polynomial continuum fit to 2.05-2.40$\mu$m for identifying the 2.20$\mu$m absorption feature in the NIMS scan.

To validate that the 2.20$\mu$m feature is real and not an artifact of noise, I investigated the continuum-removed average spectra along with their associated standard errors against the continuum baseline. The lower panel of Fig. \ref{fig:fig2} shows that, for detected pixels, the reflectance values at 2.19, 2.20, and 2.21$\mu$m—along with their standard errors—fall clearly below the continuum baseline, confirming that the feature exceeds the noise level. This observation verifies that the absorption feature at $\sim$2.20$\mu$m for the detected pixels is real and rise above the noise in each spectrum. In contrast, for non-detected pixels, the continuum baseline lies within the range of reflectance ± standard errors for these channels, indicating the absence of a true absorption feature (lower panel of Fig. \ref{fig:fig2}).

\begin{figure*}[ht!]
\plotone{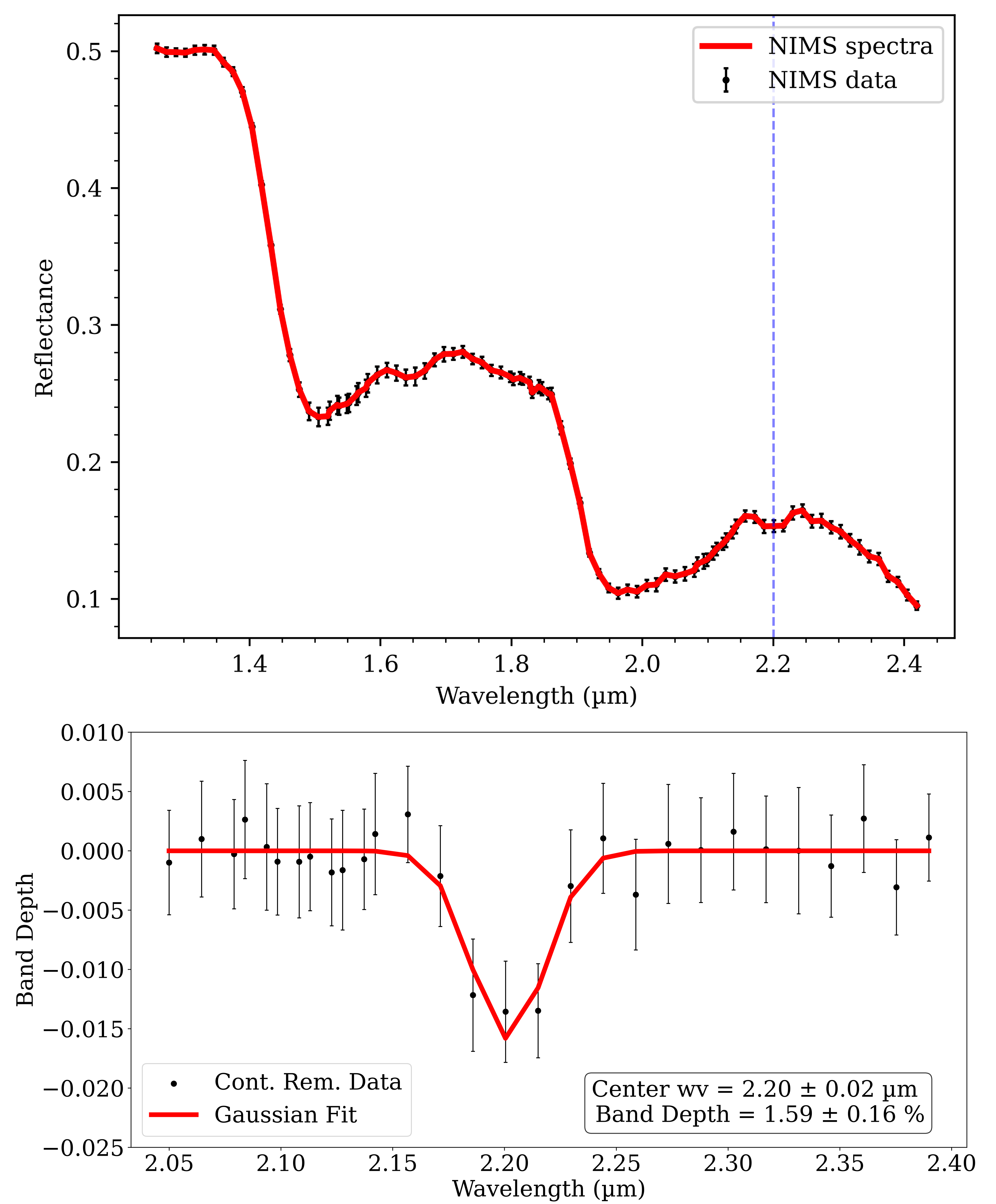}
\caption{\textit{Upper panel}: Average reflectance spectrum with corresponding standard error bars across all wavelengths for the pixels within the detected clusters from NIMS observation 11ENCYCLOD01A (cf. Fig. \ref{fig:fig4}) The broad absorption features near $\sim$1.5 and $\sim$2.0$\mu$m are characteristic of H\textsubscript{2}O ice \citep{grundy1998temperature, mastrapa2008optical}. A distinct absorption feature centered at $\sim$2.20$\mu$m– indicative of NH\textsubscript{3}-bearing species (marked by the dashed blue vertical line). Note that the average reflectance of each cluster (4 in total; cf Fig. \ref{fig:fig4}) is given in Fig. \ref{fig:fig10}. \textit{Bottom panel:} Gaussian band fit to the continuum-subtracted average spectrum shows a 2.20$\pm$0.02$\mu$m absorption feature with a band depth of 1.59$\pm$0.16\%. 
\label{fig:fig3}}
\end{figure*}

\begin{figure*}[ht!]
\plotone{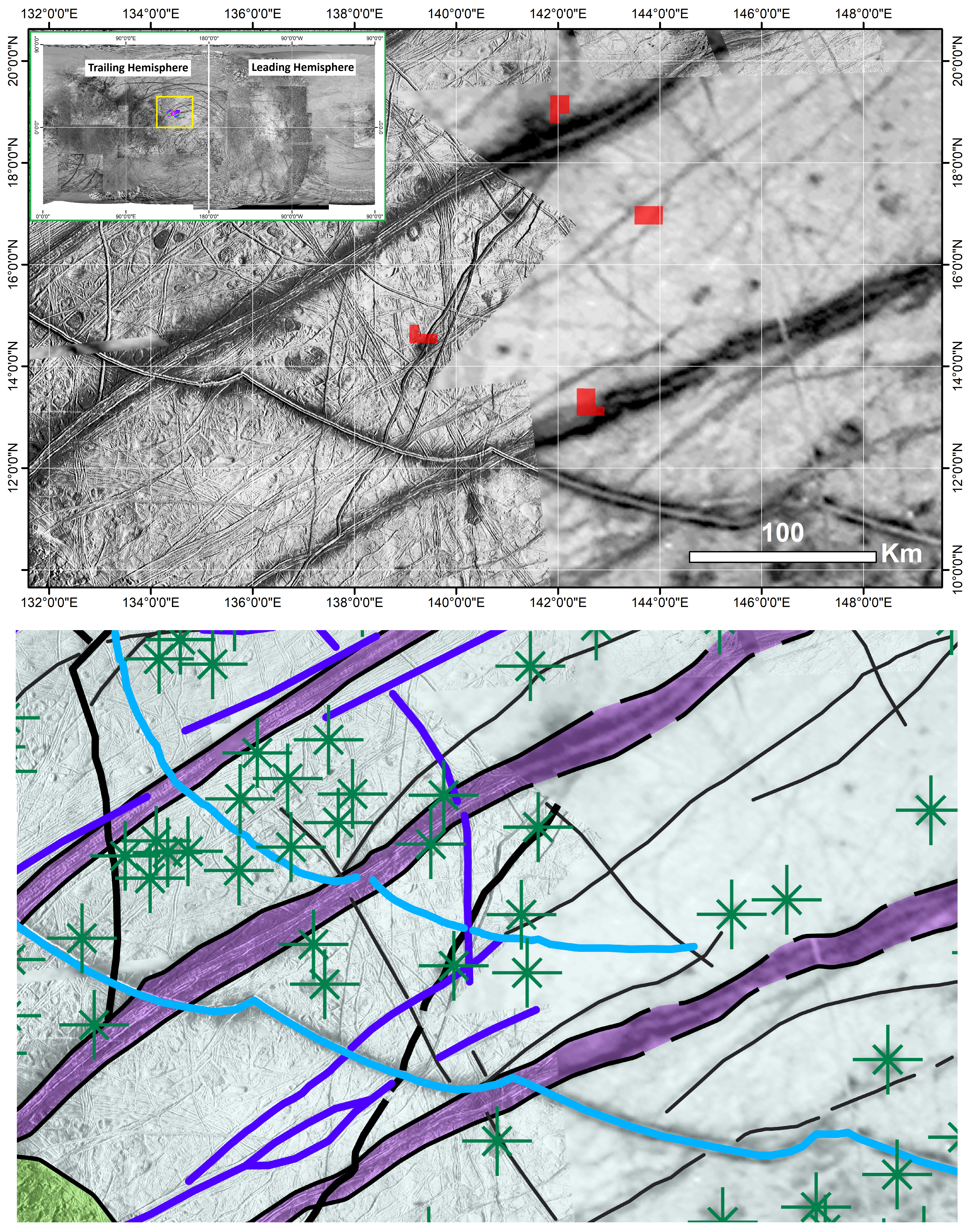}
\caption{\textit{Upper panel}: Distribution of detected pixel clusters (red polygons) overlaid on the Galileo/SSI basemap \citep[adopted from ][]{malaska2018europa, malaska2024updated}. \textit{Inset}: The yellow box shows the locational context of the investigated image on the USGS's Voyager and Galileo/SSI Global Mosaic at a resolution of 500 m/pixel\footnote{ The United States Geological Survey (USGS)’s Voyager and Galileo/SSI global mosaic can be accessed 
at \url{https://astrogeology.usgs.gov/search/map/europa_voyager_galileo_ssi_global_mosaic_500m} 
}. Note that the average spectrum of each pixel cluster is provided in Fig. \ref{fig:fig10}. \textit{Bottom panel}: A portion of the geologic map of the same region on Europa \citep[adopted from ][]{leonard2024global}. The violet color shaded feature represents the band unit, while linear features include cycloid (aqua blue line), linear band (navy blue line), depression margin (wider black line), and ridges/troughs (narrow black lines). The dark green asterisk symbol represents microchaos features. Refer to \cite{leonard2024global} for details of the map elements. The detected pixel clusters in the upper panel correlate with microchaos, linear, and band geologic units in the bottom panel.
\label{fig:fig4}}
\end{figure*}

Upon verifying the detected pixels, I applied a Gaussian fit to the continuum-removed absorption feature in the average spectrum (Fig. \ref{fig:fig3}). Note that the average reflectance and associated standard errors for the detected pixels are provided in Table \ref{tab:table2}. For reference, the average reflectance of each individual verified cluster (4 in total; cf Fig. \ref{fig:fig4}) is given in Fig. \ref{fig:fig10}. Lastly, the spatial (geographic) distribution of the detected pixels was mapped onto a high-resolution Solid State Imager experiment \citep[SSI;][]{belton1992galileo} basemap of Europa \citep{malaska2024updated} and correlate with the geologic map of Europa \citep[Fig. \ref{fig:fig4}; ][]{leonard2024global}.

\begin{figure*}[ht!]
\plotone{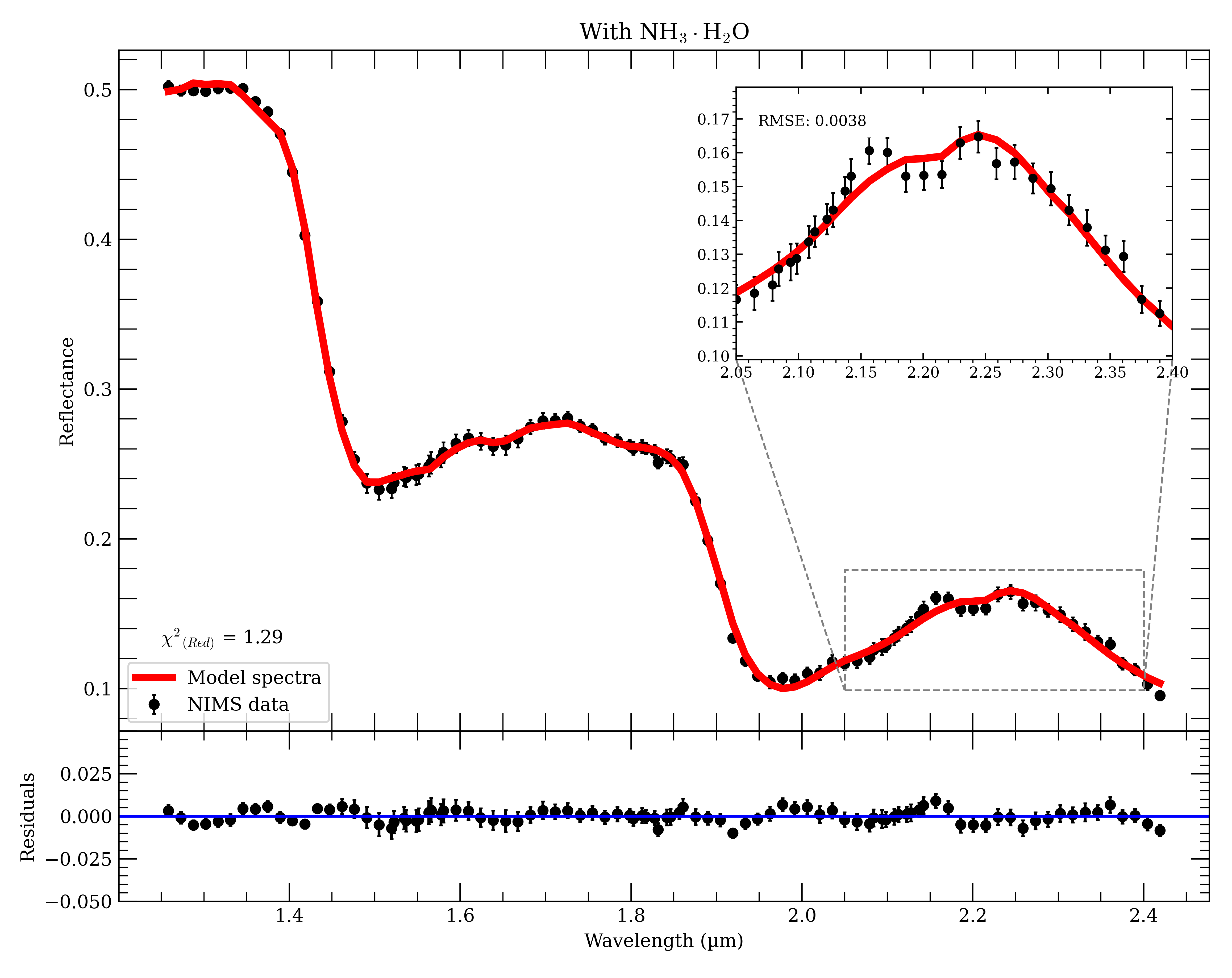}
\caption{\textit{Top panel}: The average reflectance spectrum of the detected clusters (black circles) with standard error bars, along with the best-fit model spectrum (red line) derived using an areal (linear) mixing approach incorporating the optical constants of NH\textsubscript{3}·H\textsubscript{2}O. The reduced chi-square value– $\chi$\textsuperscript{2}\textsubscript{(\textit{Red.})}– of the model fit is also provided in the plot. The inset shows a zoomed-in view of the model fit between 2.05–2.40$\mu$m, with the corresponding RMSE value. The model spectrum closely matches the observed absorption feature at 2.20$\mu$m, rendering a lower RMSE than the fit shown in Fig. \ref{fig:fig6}. \textit{Bottom panel}: A plot of the residual (observation-model) spectrum over the entire wavelength region at $\sim$1.2 - 2.5$\mu$m.
\label{fig:fig5}}
\end{figure*}

\begin{figure*}[ht!]
\plotone{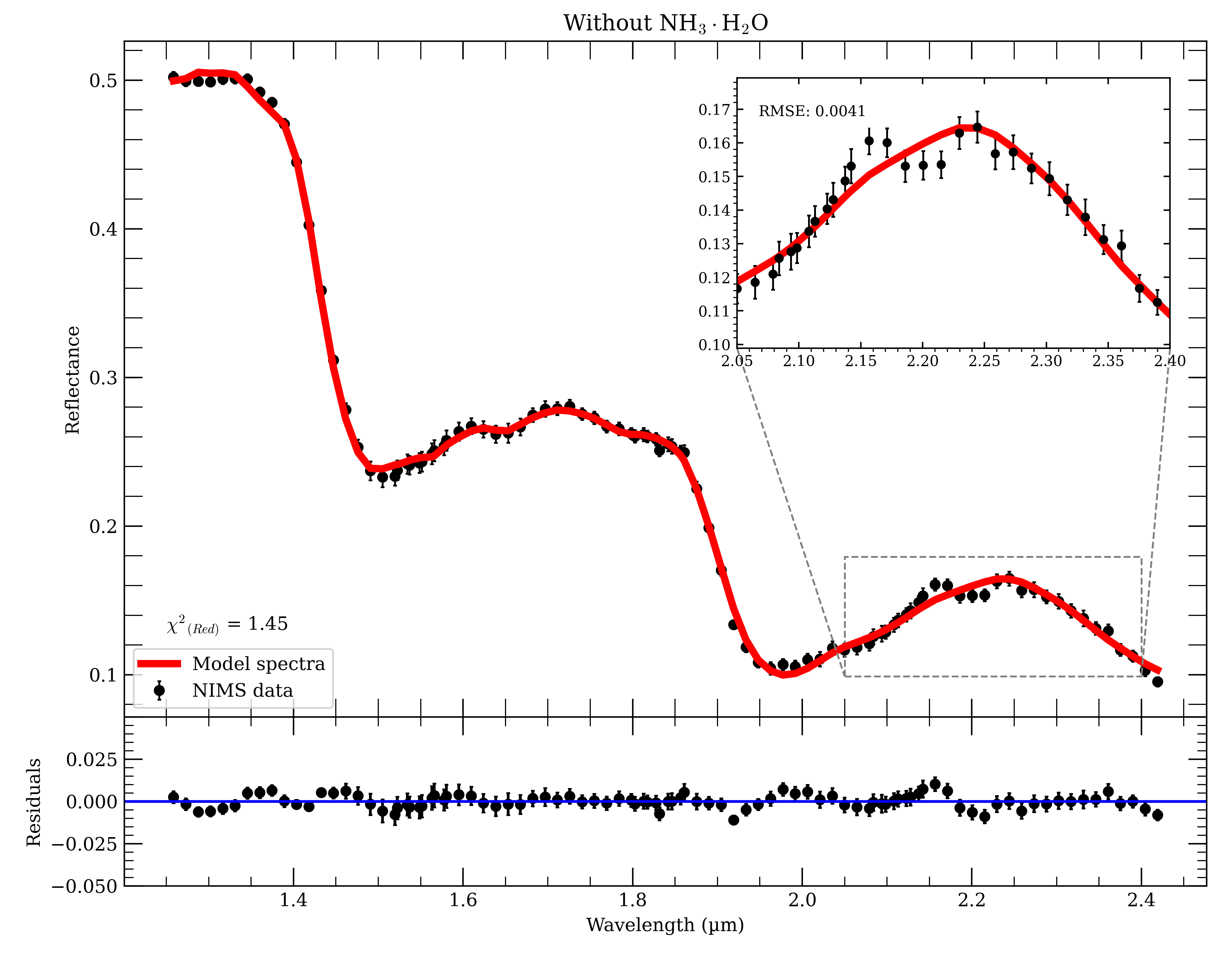}
\caption{\textit{Top panel}: The average reflectance spectrum of the detected clusters (black circles) with standard error bars, along with the best-fit model spectrum (red line) derived using an areal (linear) mixing approach that excludes the optical constants of NH\textsubscript{3}·H\textsubscript{2}O. The reduced chi-square value– $\chi$\textsuperscript{2}\textsubscript{(\textit{Red.})}– of the model fit is also provided in the plot. The inset shows a zoomed-in view of the model fit between 2.05–2.40$\mu$m, with the corresponding RMSE value. The model spectrum fails to reproduce the absorption feature at 2.20$\mu$m seen in the NIMS data and results in a higher RMSE than the fit shown in Fig. \ref{fig:fig5}. \textit{Bottom panel}: A plot of the residual (observation-model) spectrum over the entire wavelength region at $\sim$1.2 - 2.5$\mu$m.
\label{fig:fig6}}
\end{figure*}

\begin{figure*}[ht!]
\plotone{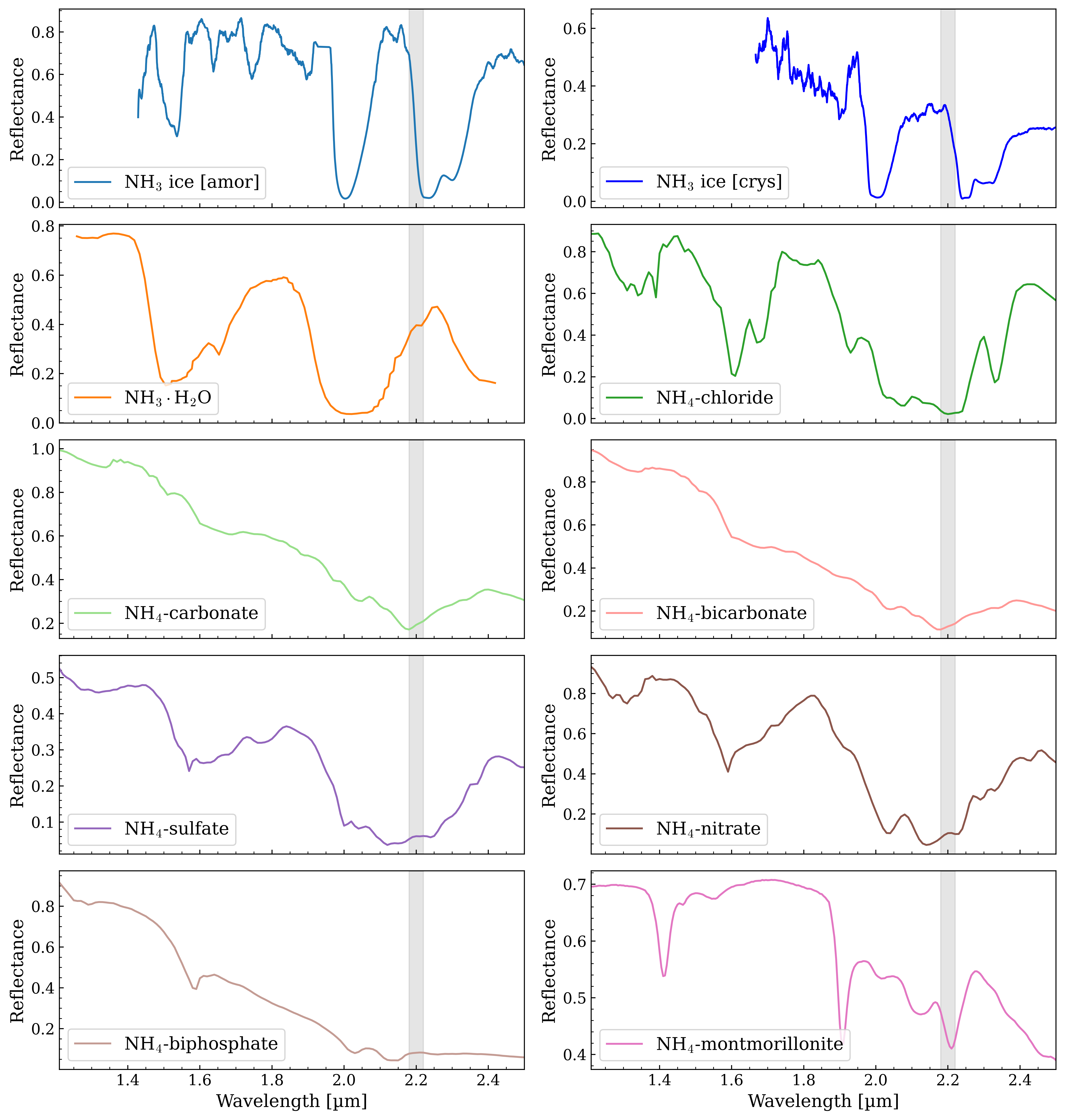}
\caption{The reflectance spectra of the NH\textsubscript{3}-bearing species at near-infrared wavelengths. The modeled reflectance spectra using the \cite{shkuratov1999model} approach for a 50$\mu$m grain for amorphous NH\textsubscript{3 }ice at 40K \citep[optical constants from][]{roser2021infrared}, crystalline NH\textsubscript{3 }ice at 100K \citep[optical constants from][]{hudson2022ammonia}, and NH\textsubscript{3}·H\textsubscript{2}O ice at 77K \citep[optical constants from][]{brown1988search}. Reflectance spectra of NH\textsubscript{4}-salts (-chloride, -carbonate, -bicarbonate, -sulfate, -nitrate, and -biphosphate) at grain sizes of 32 – 80$\mu$m measured at 120K were obtained from \cite{fastelli2022reflectance}. The spectrum of NH\textsubscript{4}-montmorillonite (a phyllosilicate; RELAB ID: c1jb189) was obtained from the RELAB spectral library \citep{milliken2021nasa}. The gray shades of the vertical area represent the wavelengths between 2.18 and 2.22$\mu$m (2.20$\pm$0.02$\mu$m), where NIMS data of the detected pixels show an absorption feature. Note that the NH\textsubscript{3} ice spectra were smoothed using the Savitzky–Golay filter \citep{savitzky1964smoothing}, following \cite{emran&Chevrier2022}. 
\label{fig:fig7}}
\end{figure*}

\begin{deluxetable*}{lccc}
\tablewidth{0pt}
\tablecaption{Best-fit parameters from linear spectral modeling, estimated from 1,000 synthetic spectra generated around the average reflectance of the detected NIMS pixel clusters, when the model was implemented including NH\textsubscript{3}·H\textsubscript{2}O. Listed are only those endmembers with modeled abundances greater than 1\%. The values represent relative areal abundances (mean$\pm$1$\sigma$) in percentage and estimated grain sizes in $\mu$m for each component. Grain sizes were estimated only for those endmembers for which optical constant data were available (see spectral modeling section in the main text)  \label{tab:table1}}
\tablehead{
\colhead{Species} & \colhead{Formula} & \colhead{Abundance (\%) } & \colhead{Diameter ($\mu$m) }}
\startdata
Water ice (a)  &  H\textsubscript{2}O (a)  & 14.54 ± 2.53 & 1757.95 ± 209.75  \\
Water ice (c)  & H\textsubscript{2}O (c) & 4.90 ± 2.48  & 1795.13 ± 256.03  \\
Hexahydrite  & MgSO\textsubscript{4}·6H\textsubscript{2}O  & 4.68 ± 0.72  & 50.9 ± 22.06  \\
Mirabilite   & Na\textsubscript{2}SO\textsubscript{4}·10H\textsubscript{2}O  & 15.91 ± 5.15  & 610.83 ± 127.26  \\
SAO  & H\textsubscript{2}SO\textsubscript{4}·8H\textsubscript{2}O  & 36.57 ± 0.42  & 28.66 ± 1.16  \\
Ammonia-hydrate  & NH\textsubscript{3}·H\textsubscript{2}O  &7.44 ± 0.63  & 93.78 ± 14.31  \\
Mg-chlorate   & Mg(ClO\textsubscript{3})\textsubscript{2}·6H\textsubscript{2}O  &4.21 ± 1.97  & --- \\
Na-perchlorate   & NaClO\textsubscript{4}  & 10.53 ± 1.29  & --- \\
RMSE  & ---  & 0.001&  --- \\
\enddata
\tablecomments{\textit{1. Water (H\textsubscript{2}O) ice (a) represents the amorphous phase and H\textsubscript{2}O (c) represents the crystalline phase. \\
2. SAO represents sulfuric acid octahydrate.\\ 
3. RMSE represents root mean square error. \\
4. Hexahydrite and mirabilite are sulfate salts. \\
5. Mg-chlorate and Na-perchlorate are Cl-bearing salts}.}
\end{deluxetable*}

\subsection{Spectral modeling} \label{subsec:model}
I applied linear spectral modeling viz. areal mixture approach \citep[e.g.,][]{daltonetal2012, emran2021thermophysical, emran&stack2025} using a radiative transfer modeling based on Shkuratov's theory \citep{shkuratov1999model} to estimate the relative abundances of ice and non-ice materials on Europa. This approach has successfully been implemented in surface compositional analyses of the moon using both NIMS data \citep{dalton2007linear, shirley2010europa, daltonetal2012, prockter2017surface, davis2024pwyll} and ground-based observations \citep{brown2013salts, ligier2016vlt}. Moreover, existing literature shows that linear and intimate mixture models render similar compositional results for Europa’s surface when applied to NIMS data \citep{shirley2016europa}. In the linear mixing model, the observed spectrum is represented as the weighted sum of the reflectance spectra of individual endmembers, with the weights corresponding to their fractional areal contributions \citep[e.g.,][]{emran2023surface, emran2025SP1}:

\[
r_{f} = \sum_{i=1}^{k} f_i \cdot r_i \quad \text{; } \quad 0 \leq f_i \leq 1
\]

 where, \textit{r\textsubscript{f}} is the NIMS reflectance of the spectrum, \textit{r\textsubscript{i}} is the reflectance of the \textit{i\textsuperscript{th}} constituent endmember, and \textit{f\textsubscript{i} }is the fraction contributed (abundance) by the \textit{i\textsuperscript{th}} endmember \citep{daltonetal2012, emran2021thermophysical, emran&stack2025, emran2025SP2}.

The candidate endmember species, including ice and non-ice components, were selected based on materials detected from ground and space-based observations or hypothesized to be present on Europa’s surface \citep[][and references therein]{davis2024pwyll, becker2024exploring}. I used optical constants for both amorphous and crystalline H\textsubscript{2}O ice at 120K from \cite{mastrapa2008optical, mastrapa2009optical}. For non-ice components, I included optical constants of sulfate salts such as hexahydrite (MgSO\textsubscript{4}·6H\textsubscript{2}O), epsomite (MgSO\textsubscript{4}·7H\textsubscript{2}O), and bloedite Na\textsubscript{2}Mg(SO\textsubscript{4})\textsubscript{2}·4H\textsubscript{2}O at 120K \citep{dalton&pitman2012}. The optical constants of mirabilite (Na\textsubscript{2}SO\textsubscript{4}·10H\textsubscript{2}O) at 100K were estimated using reflectance spectra measured at various grain sizes from \cite{de2021temperature}. This was done based on the Shkuratov theory \citep{shkuratov1999model}, following the approach of \cite{mermy2023selection}. Sulfuric acid octahydrate (SAO; H\textsubscript{2}SO\textsubscript{4}·8H\textsubscript{2}O) optical constants at 77K were obtained from \cite{carlson2005distribution}.

Due to the lack of comprehensive optical constants for NH\textsubscript{3}-bearing species, I used the optical constants of NH\textsubscript{3}·H\textsubscript{2}O ice at 77K from from \citealt[][published in \citealt{cruikshank2005spectroscopic}]{brown1988search}, as a proxy \citep[e.g.,][]{emran2023surface, decolibus2023nh3}. For chlorine (Cl-bearing) salts, where optical constants are unavailable or grain size information is missing, I used reflectance spectra at 80K from \cite{hanley2014reflectance} for magnesium chlorate (Mg(ClO\textsubscript{3})\textsubscript{2}·6H\textsubscript{2}O), magnesium perchlorate (Mg(ClO\textsubscript{4})\textsubscript{2}·6H\textsubscript{2}O), magnesium chloride (MgCl\textsubscript{2}·2H\textsubscript{2}O; MgCl\textsubscript{2}·4H\textsubscript{2}O; and MgCl\textsubscript{2}·6H\textsubscript{2}O), sodium chloride (NaCl), and sodium perchlorate (NaClO\textsubscript{4} and NaClO\textsubscript{4}·2H\textsubscript{2}O). The examples of reflectance spectra of all endmembers used in the linear model were given in Fig. \ref{fig:fig11}. I estimated grain diameters for endmembers with optical constants by fitting the model spectra. The mean and standard errors of the best-fit parameters were estimated from 1,000 randomly generated synthetic reflectance spectra, created by adding noise to the average spectrum based on its standard error \citep{emran2021thermophysical, emran&stack2025}. The estimated fractional abundances and grain sizes from the spectral model are listed in Table \ref{tab:table1}, and the best-fit model spectrum over the wavelengths is shown in Fig. \ref{fig:fig5}.

\section{Results} \label{sec:results}
The average spectrum of the detected pixels (Fig. \ref{fig:fig3}) exhibits broad absorption features near $\sim$1.5 and $\sim$2.0$\mu$m, characteristic of H\textsubscript{2}O ice \citep[e.g.,][]{mastrapa2008optical, mastrapa2009optical, emran&Chevrier2023}. Additionally, an absorption feature near 1.65$\mu$m may indicate the presence of crystalline H\textsubscript{2}O ice or a mixture of amorphous and crystalline phases \citep{grundy1998temperature, mastrapa2008optical}. A distinct absorption feature centered at $\sim$2.20$\mu$m indicates the presence of NH\textsubscript{3}-bearing species \citep{moore2007ammonia, carlson2009europa, roser2021infrared, fastelli2020reflectance, fastelli2022reflectance}. The Gaussian band fitting of the continuum-subtracted data results in a central wavelength of 2.20 ± 0.02$\mu$m and a band depth of 1.59 ± 0.16\% (Fig. \ref{fig:fig3}). The corresponding pixels are spatially associated with micro chaos, linear, and band geologic units on the trailing hemisphere (Fig. \ref{fig:fig4}). 

Previous spectroscopic investigations of NIMS data have reported the presence of hydrated sulfate salts, including sodium sulfates (Na\textsubscript{2}SO\textsubscript{4}·\textit{n}H\textsubscript{2}O) and magnesium sulfates (MgSO\textsubscript{4}·\textit{n}H\textsubscript{2}O), on Europa \citep[e.g.,][]{mccord1998salts, mccord2002brines, carlson2009europa}. While these hydrated sulfates salts may show a weak absorption near $\sim$2.2$\mu$m, the feature is inherently broad and asymmetric, spanning $\sim$2.18–2.24$\mu$m \citep[cf. Fig. \ref{fig:fig11}; e.g.,][]{mccord2002brines, carlson2009europa, dalton&pitman2012, de2021temperature}. In particular, sodium sulfates— such as mirabilite \citep[Na\textsubscript{2}SO\textsubscript{4}·10H\textsubscript{2}O; ][]{de2021temperature} and frozen Na\textsubscript{2}SO\textsubscript{4} brine \citep[e.g.,][]{mccord2002brines} — along with other hydrated sulfates, show a strong absorption band near $\sim$1.78$\mu$m \citep[e.g.,][]{carlson2009europa} relative to the weaker $\sim$2.2$\mu$m feature (Fig.\ref{fig:fig11}). Thus, if the $\sim$2.2$\mu$m feature were due to Na\textsubscript{2}SO\textsubscript{4} salts, the corresponding NIMS spectra should also exhibit a stronger $\sim$1.78$\mu$m absorption—an observation that that is inconsistent the NIMS spectra investigated here. In contrast, NH\textsubscript{3}-bearing compounds, in general, show a relatively sharp absorption band at $\sim$2.20$\mu$m \citep[e.g.,][]{moore2007ammonia, bergetal2016, fastelli2022reflectance}, which closely resembles the feature detected in the NIMS data (2.20 ± 0.02$\mu$m; Fig. \ref{fig:fig3}). Thus, this observation, a $\sim$2.20$\mu$m feature, suggests the presence of NH\textsubscript{3}-bearing species— such as NH\textsubscript{3}-hydrate \citep[e.g.,][]{carlson2009europa}. Nonetheless, a mixture of minerals— such as NH\textsubscript{3}-bearing components with hydrated sulfate salts— is plausible and further explored in the spectral modeling results presented below. 

Some metal-OH clays— such as Fe-, Mg-, and Al-rich phyllosilicate minerals—has been hypothesized to present on Europa \citep[e.g.,][]{mccord1998salts, carlson2009europa} and are known to exhibit absorption features within the 2.2–2.4$\mu$m range \citep[Fig. \ref{fig:fig12}; e.g.,][]{hunt1970visible, clark1990high, clark2020spectroscopy}. However, such absorption features have not been confidently reported on Europa’s spectra using NIMS observations \citep[e.g.,][]{mccord1998salts}. While these phyllosilicate minerals containing interlayer H\textsubscript{2}O (i.e., water in clay) can distort and show asymmetric water-related bands \citep{clark1990high, clark1993us}, the spectral shapes and band combinations— such as H-O-H stretching plus bending at 2$\mu$m— are inconsistent with the spectral characteristics observed in Europa's NIMS data at 1-3$\mu$m region \citep[e.g.,][]{hunt1979spectra, clark1990high, mccord1998salts, carlson2009europa}. Moreover, hydrated phyllosilicates often show narrower and symmetric absorption features at shorter wavelengths \citep[e.g.,][]{mccord1998salts}— such as near $\sim$1.39$\mu$m \citep{clark1990high}— which are inconsistent with the NIMS data analyzed here (Fig. \ref{fig:fig12}). Thus, the absorption feature detected at $\sim$2.20$\mu$m in this study is unlikely to originate from clay-bearing (metal OH-bearing phyllosilicate) minerals. 

To further validate a contribution of NH\textsubscript{3}-bearing compounds in the NIMS spectra, I applied the linear spectral modeling with and without the inclusion of NH\textsubscript{3}·H\textsubscript{2}O, keeping all other endmembers constant (Figs. \ref{fig:fig5}-\ref{fig:fig6}). The resulting reduced chi-square - $\chi$\textsuperscript{2}\textsubscript{(\textit{Red.})} - values were 1.29 and 1.45, respectively, for the fits with and without NH\textsubscript{3}·H\textsubscript{2}O to the model. Furthermore, the model including NH\textsubscript{3}·H\textsubscript{2}O more closely reproduces the observed absorption feature at $\sim$2.20$\mu$m, resulting in a lower root mean square error (RMSE) in the 2.05–2.40$\mu$m region (insets of Figs. \ref{fig:fig5}-\ref{fig:fig6}). This supports the presence of an ammoniated component in the NIMS spectra— indicating that hydrated sulfate salts (Na\textsubscript{2}SO\textsubscript{4}·\textit{n}H\textsubscript{2}O and MgSO\textsubscript{4}·\textit{n}H\textsubscript{2}O) alone cannot explain the observed spectra of the detected pixels. To assess whether the improvement in model fit due to the inclusion of NH\textsubscript{3}·H\textsubscript{2}O is statistically significant, I performed an F-test \citep[e.g.,][]{bevington2003data} comparing the two models \citep[e.g.,][]{emran&stack2025}. The resulting \textit{p}-value (0.0145) indicates that the inclusion of NH\textsubscript{3}·H\textsubscript{2}O significantly improves (better model) the fit at the 95\% confidence interval. These results evidently support the contribution of ammoniated compounds to the observed spectral signature on Europa— NH\textsubscript{3}-bearing components are required to explain the NIMS observation. Thus, the spectral modeling results presented and described in this paper were obtained by incorporating NH\textsubscript{3}·H\textsubscript{2}O into the model.

Spectral modeling results, with incorporating NH\textsubscript{3}·H\textsubscript{2}O (Table \ref{tab:table1}), show the highest abundance ($\sim$37\%) of sulfuric acid octahydrate (H\textsubscript{2}SO\textsubscript{4}·8H\textsubscript{2}O), consistent with expectations for the trailing hemisphere, where irradiation by Jupiter’s magnetosphere resulted in radiolytically alteration of sulfur to sulfuric acid hydrate \citep{carlson1999sulfuric, carlson2005distribution, daltonetal2012, brown2013salts, fischer2015spatially}. On the trailing hemisphere, the exogenic sulfur is hypothesized to be implanted from Jupiter’s moon Io \citep{carlson1999sulfuric, carlson2002sulfuric, carlson2005distribution}. An abundance of $\sim$40-50\% of SAO at the chaos terrains in the trailing hemisphere of the moon has been suggested by previous spectral analysis \citep{daltonetal2012, fischer2015spatially, ligier2016vlt}. NH\textsubscript{3}-bearing species (ammonia-hydrate in this instance) constitute approximately 7\% of the modeled composition. Using spectral analysis, \cite{emran2023surface} suggested the presence of $\sim$15\% ammoniated content at a cryovolcanic caldera site on Pluto \citep{emran2025kiladze}. 

The sulfate salts (hexahydrite and mirabilite; $\sim$21\%) are found at higher abundance than the abundance of Cl-bearing salts (Mg-chlorate and Na-perchlorate; $\sim$15\%). A combination of hexahydrite and mirabilite, representing the major sulfate salts, has been previously identified on Europa’s bright plains and dark linea using NIMS observations \citep{daltonetal2012}. The presence of chlorine salts on Europa has been confirmed by a ground-based Subaru/IRCS telescope with a conservative upper limit of 17\% \citep{tan2022spatially}. Using the observation from the Very Large Telescope (VLT)’s near-infrared instrument,  \cite{ligier2016vlt} suggest an abundance of hydrated Cl-bearing salts to be \(>20\% \) in the regions in the trailing hemisphere. Although the non-ice salt grain sizes are generally below 100$\mu$m (except mirabilite), the modeled H\textsubscript{2}O ice (both amorphous and crystalline phases) grains are notably coarser, \( >1 \)mm. However, amorphous H\textsubscript{2}O ice ($\sim$15\%) is $\sim$3x times more abundant than crystalline H\textsubscript{2}O ice ($\sim$5\%). While the global distribution of crystalline H\textsubscript{2}O ice correlates with geomorphological units on Europa, its average abundance remains \( <15\% \) across the lower latitudes in the trailing hemisphere \citep{ligier2016vlt}. \cite{ligier2016vlt} also reported that crystalline H\textsubscript{2}O ice with grain sizes of $\sim$1 mm exclusively on the trailing hemisphere, particularly along large lineae \citep{doggett2009geologic, leonard2024global}. 

While all other non-ice components showed grain sizes smaller than 100$\mu$m, the sulfate salt mirabilite (Na\textsubscript{2}SO\textsubscript{4}·10H\textsubscript{2}O) showed comparatively larger grains ($\sim$600$\mu$m). This may reflect a real compositional scenario in which mirabilite grains are inherently larger, or an “unrealistic” model parameter that could result from the use of calculated optical constants derived from reflectance spectra at different grain sizes \citep{de2021temperature}, and therefore should be interpreted with caution. Nonetheless, the estimated abundance of mirabilite is consistent with values reported elsewhere on Europa by using both NIMS and ground-based observations \citep{dalton2007linear, daltonetal2012, brown2013salts}. Laboratory measurements of optical constants for mirabilite, Cl-bearing salts, and ammoniated components at Europa’s relevant temperature are warranted to improve the accuracy of compositional interpretations.

\section{Discussion} \label{sec:discussion}

I have detected a distinct absorption feature at 2.20 ± 0.02$\mu$m, suggesting the presence of ammoniated components such as hydrates, salts, and phyllosilicates \citep{moore2007ammonia, roser2021infrared, fastelli2022reflectance}, on Europa using Galileo/NIMS data. NH\textsubscript{3} ice (both amorphous and crystalline) has an absorption band shifted toward a longer wavelength near 2.24$\mu$m \citep[Fig. \ref{fig:fig7};][]{roser2021infrared, hudson2022ammonia}. Moreover, pure NH\textsubscript{3} is also highly unstable in the space environment from solar wind, solar UV particles, and Galactic cosmic rays \citep{moore2007ammonia, cruikshank2019b} compared to the other NH\textsubscript{3}-bearing compounds \citep{altwegg2020evidence, decolibus2023nh3, nakazawa2025nitrogen}. In contrast, the absorption band of NH\textsubscript{3}·H\textsubscript{2}O typically centers around $\sim$2.21$\mu$m, but the exact position varies with the NH\textsubscript{3}:H\textsubscript{2}O mixing ratio—shifting from 2.229$\mu$m to 2.208$\mu$m as NH\textsubscript{3} content decreases from 100\% to 1\% \citep{zheng2009infrared}. Among the ammonia-bearing salts, NH\textsubscript{4}-chloride (NH\textsubscript{4}Cl) has an absorption feature near 2.20$\mu$m \citep{bergetal2016, fastelli2022reflectance}, closely matching the detection here (cf. Figs. \ref{fig:fig3} and \ref{fig:fig7}). Other NH\textsubscript{3}-bearing salts, such as NH\textsubscript{4}-carbonate, NH\textsubscript{4}-bicarbonate, NH\textsubscript{4}-sulfate, NH\textsubscript{4}-nitrite, and NH\textsubscript{4}-phosphate, have band centers in the shorter wavelength region $\sim$ 2.17-2.15$\mu$m \citep[Fig. \ref{fig:fig7}:][]{bergetal2016, fastelli2022reflectance}. Although NH\textsubscript{4}-phyllosilicate (a hydrated silicate) has an absorption feature near $\sim$2.20$\mu$m (Fig. \ref{fig:fig7}), this is not its primary band and is therefore unlikely to appear in a mixed spectrum \citep[e.g.,][]{cartwright2023evidence, decolibus2023nh3}. Furthermore, as mentioned earlier, hydrated silicates have a stronger absorption near $\sim$1.39$\mu$m \citep{clark1990high}, along with other narrow and symmetric features at shorter wavelengths \citep[e.g.,][]{mccord1998salts}, which are unrecognizable or not evident in the Galileo/NIMS spectrum (cf. Fig. \ref{fig:fig3} and \ref{fig:fig12}). The spectral modeling (Table \ref{tab:table1}) indicates the presence of NH\textsubscript{3}·H\textsubscript{2}O and Cl-bearing salts. Thus, based on the band position (Fig. \ref{fig:fig3}) and spectral modeling (Table \ref{tab:table1}), I suggest that NH\textsubscript{3}-hydrate and NH\textsubscript{4}-chloride are the most likely contributors to the observed 2.20$\mu$m feature. However, the limited spectral resolution of the Galileo/NIMS observation, coupled with the lack of comprehensive laboratory optical constant data for NH\textsubscript{3}– species, does not rule out the presence of other NH\textsubscript{3}-bearing components on Europa. Future investigations using higher resolution spectral data from ESA's JUICE and NASA's Europa Clipper missions \citep{pappalardo2024science} will help to understand the specific NH\textsubscript{3}-bearing species on the moon.

 The spatial correlation between the detected pixels and microchaos, linear, and band geologic units suggests that H\textsubscript{2}O carrying NH\textsubscript{3}-bearing species was emplaced onto Europa’s surface via cryovolcanic or similar processes from the subsurface or underground ocean. The presence of chlorine-bearing materials in Europa’s subsurface ocean has been previously proposed \citep{kargel2000europa, zolotov2001composition, trumbo2019sodium}. Thus, it is plausible that NH\textsubscript{3}-rich liquids and Cl-bearing brines– such as NH\textsubscript{3}·H\textsubscript{2}O and NH\textsubscript{4}Cl– may have been emplaced onto the surface via cryovolcanic processes, as these materials \citep{emran2023surface, cartwright2020evidence, cruikshank2019b, cook2007near} and associated geologic features are considered manifestations of cryovolcanism and surface–subsurface material exchange \citep{wilson1997eruption, pappalardo1998geological, pappalardo1999does, daubar2024planned}. Particularly, the microchaos terrain– referred to as “pits, domes, or spots”, and also known as lenticulae in the literature– is the youngest geologic feature on Europa \citep[e.g.,][]{figueredo2004resurfacing, prockter1999europa, doggett2009geologic, leonard2018analysis, leonard2024global}. The widespread presence of chaos terrains suggests that Europa’s subsurface ocean is close to the surface, with the possibility of subsurface lakes within the ice shell or forming in the regions where the ice shell thickness approaches zero \citep{greenberg1999chaos, schmidt2011active}. These chaos terrains are thought to develop through processes such as ice shell convection, mobilization of brines, or the eruption of cryolava from liquid reservoirs \citep[refer to ][for exhaustive references]{daubar2024planned}. Therefore, chaos terrains are closely associated with and an indicator of surface–subsurface interactions and/or cryovolcanic activity \citep[e.g.,][]{howell2024jupiter}. On the other hand, the band and linear features are likely tectonic in origin, such that bands are associated with a similar process of rifting and spreading at mid-ocean ridges  \citep{greeley2000geologic, howell2018band, daubar2024planned, leonard2024global}. The linear and band units might have acted as conduits through which ammoniated components were transported from the underground ocean or shallow subsurface and emplaced onto the surface– a similar emplacement mechanism for the ammoniated contents in Virgil Fossae and Viking Terra on Pluto \citep{dalle2019detection, cruikshank2019b}. Thus, the geologic context of the detected ammoniated material suggests that Europa’s subsurface ocean or briny liquid reservoirs may be accessible at shallow depths.
 
The presence of sulfate salts (Table \ref{tab:table1}) may result from post-emplacement chemical alteration, where Cl-bearing salts from the underground ocean \citep{kargel2000europa, zolotov2001composition} may be converted to sulfate salts upon exposure to the surface (Jupiter’s radiation) as an irradiation product \citep{brown2013salts, hand2015europa}. However, an emplacement of sulfate salts from the ocean has also been proposed \citep{mccord1999hydrated, mccord2010hydrated, Zolotov2009}. While mirabilite (Na\textsubscript{2}SO\textsubscript{4}·10H\textsubscript{2}O) appears to be more abundant than hexahydrite (MgSO\textsubscript{4}·6H\textsubscript{2}O) in this analysis, hexahydrite is more stable than mirabilite under Europa’s radiation environment \citep{johnson2000sodium, johnson2001surface, mccord2001thermal, daltonetal2012}. Detection of a weak feature near 2.07$\mu$m in ground-based observations (W. M. Keck Observatory) has previously been attributed to a possible Mg-sulfate salt or brine, likely as an irradiation product of endogenic material originating from Europa’s subsurface and exposed on the trailing hemisphere \citep{brown2013salts}. However, an alternative explanation for this feature—such as a product of the radiolytic sulfur cycle or an unidentified material resulting from parallel irradiation processes—has also been proposed \citep{davis2023spatial}. While the average spectrum in this study may exhibit a weak absorption near 2.07$\mu$m (Fig. \ref{fig:fig3}), the subtlety of this feature is likely due to the limited spectral resolution of the Galileo/NIMS dataset. Nevertheless, further investigation—particularly focused on confirming the 2.07$\mu$m band in the NIMS data—is warranted. Higher-resolution imaging spectrometers, such as JUICE’s Moons and Jupiter Imaging Spectrometer \citep[MAJIS;][]{poulet2024moons} and Europa Clipper’s Mapping Imaging Spectrometer for Europa \citep[MISE:][]{blaney2024mapping}, are expected to help resolve this feature on the moon with higher confidence.

Spectral modeling (Table \ref{tab:table1}) shows relatively large grain sizes for H\textsubscript{2}O ice, which may be indicative of reduced porosity and cementation within the regolith due to the presence of ammonia hydrates \citep{ahrens2020}. On Earth, ice sheets under conditions similar to those at the base of Europa’s ice shell typically consist of grain sizes on the order of $\sim$1–5$\mu$m \citep{deLaChapelle1998dynamic, barr2005onset}. Notably, the ice shells on other icy bodies in the outer solar system, such as Pluto, may have non-Newtonian rheology due to larger grains \citep{kamata2019pluto}. A similar rheological regime may exist within Europa’s ice shell. Large grain sizes ($\sim$1mm) around large lineae in the trailing hemisphere have been reported by an existing study \citep{ligier2016vlt}. The large grain sizes estimated here are consistent with effusive cryovolcanic emplacement (or a similar process such as surface flow) rather than explosive activity. This mechanism contrasts with Enceladus, where plume materials contain micron- to submicron-sized grains, indicative of explosive eruptions \citep{kempf2018saturn, postberg2018plume, klenner2024identify}. Note that no evidence of active plume activity on Europa was found using the James Webb Space Telescope (JWST) observation \citep{villanueva2023endogenous}, though the possibility of plume activity cannot be ruled out \citep{daubar2024planned}. Nonetheless, the presence of ammoniated content suggests that Europa’s ice shell is a few tens of kilometers thinner —and correspondingly thicker, reduced high-pH subsurface ocean \citep{handetal2009}— than it would be if composed solely of H\textsubscript{2}O ice \citep{spohn&schubert2003}. Heat flux models predict an ice-shell thickness ranging from $\sim$23 to 47 km \citep{howell2021likely}, although estimates up to 90 km have also been proposed \citep[e.g.,][]{vilella2020tidally}.

A recent JWST NIRSpec study reported evidence of recent surface modification on Europa, indicated by the detection of crystalline H\textsubscript{2}O ice on the surface of the moon \citep{cartwright2025jwst}. Since crystalline H\textsubscript{2}O ice can be removed rapidly via energetic particle (irradiation) driven amorphization —on timescales of less than 15 days within the top $\sim$10$\mu$m of Europa’s regolith at lower latitudes — its presence suggests ongoing or recent resurfacing processes and a vertically stratified regolith on the moon \citep{cartwright2025jwst}. The wavelength range of the NIMS data used in this study is not optimal for resolving the crystalline H\textsubscript{2}O ice Fresnel peak near 3.1$\mu$m. However, the presence of the 1.65$\mu$m band and the spectral modeling confirm the presence of crystalline H\textsubscript{2}O ice, although at an abundance much lower than the amorphous phase (refer to Table \ref{tab:table1}). Nonetheless, the relative abundances support that the materials in the studied region have not been exposed long enough to undergo complete space weathering, which would amorphize H\textsubscript{2}O ice, completely remove NH\textsubscript{3}-bearing species, and likely convert chlorine salts into more stable sulfate salts as irradiation products. On Europa’s trailing hemisphere, the equilibrium timescale for irradiation-induced alteration (sulfur cycle) is estimated to be $\sim$10\textsuperscript{4} years \citep{carlson2002sulfuric, ding2013implantation}. Whereas ammoniated species are expected to be removed from the surface within \(<10\)\textsuperscript{6} years in the space environment \citep{moore2007ammonia}, particularly Europa being exposed to Jupiter’s intense magnetospheric charged particles \citep{paranicas2001electron, cooper2001energetic}. Although the exact emplacement age of materials in this region cannot be precisely constrained, the preservation of ammoniated components indicates that their deposition is geologically recent in the context of astronomical timescales \citep{cartwright2020evidence, dalle2019detection, emran2023surface, emran2025kiladze}, given that ammonia has a short lifespan in the space environment \citep{loeffler2006enceladus, moore2007ammonia}.

While this study focused on a small region reporting the first detection of NH\textsubscript{3}-bearing material on Europa, extending the analysis to other areas using available Galileo/NIMS data is warranted. Investigations of other regions would likely reveal ammoniated materials associated with similar geologic units, assuming the compounds were emplaced and survived on the surface at the time of the NIMS observations. However, detections elsewhere on Europa are constrained by the sparse spatial coverage and variable quality of the NIMS dataset. Nonetheless, a broader survey would help to further constrain the global distribution of NH\textsubscript{3}-bearing compounds, their associations with geologic terrains, and the relative timescales of emplacement. These questions could be directly addressed by upcoming high-resolution datasets from the Europa Clipper \citep{pappalardo2024science}, which will enable a more systematic search for ammoniated species across Europa’s surface.

\section{Conclusion} \label{sec:conclusion}
In this study, I report the detection of a 2.20$\mu$m absorption feature on Europa using reprocessed Galileo/NIMS data. Spectral analysis and linear modeling suggest that the most plausible carriers of this feature are NH\textsubscript{3}-hydrate and NH\textsubscript{4}-chloride. The spatial distribution of the detected pixels is consistent with geologically young terrains, such as microchaos, linear, and band geologic units, supporting the hypothesis that NH\textsubscript{3}-bearing materials were emplaced via effusive cryovolcanism or a similar mechanism (endogenic process) involving transport of materials from a subsurface ocean or localized liquid reservoirs. This research provides evidence for recent geological activity in Europa and supports the possibility of material exchange between the surface and the underground ocean or subsurface reservoirs. The transport of NH\textsubscript{3}-bearing material from subsurface sources provides insight into the composition and chemistry of Europa’s interior, suggesting a chemically reduced high-pH \citep{handetal2009} and thicker subsurface ocean beneath a comparatively thinner ice shell \citep{spohn&schubert2003}. Nonetheless, the detection of NH\textsubscript{3}-bearing components in this study provides the first evidence of nitrogen-bearing species on Europa, an observation of considerable astrobiological significance due to nitrogen’s foundational role in the molecular basis of life.

\section*{Data Availability } \label{sec:dataavai}
All data used in this study can be found in the National Aeronautics and Space Administration's Planetary Data System: Imaging Node Server at \href{https://pdsimage2.wr.usgs.gov/}{https://pdsimage2.wr.usgs.gov/} \citep{malaska2024updated}. The optical constants data are collected from the cited references. Laboratory spectral data can be accessed from SSHADE at \href{https://www.sshade.eu/}{https://www.sshade.eu/} \citep{schmitt2018solid} and RELAB at \href{https://pds-speclib.rsl.wustl.edu/}{https://pds-speclib.rsl.wustl.edu/} \citep{milliken2021nasa}. The average reflectance and associated standard error data from all detected pixels are provided in Table \ref{tab:table2}.

\begin{acknowledgments}
This research was carried out at the Jet Propulsion Laboratory (JPL), California Institute of Technology, under a contract with the National Aeronautics and Space Administration (80NM0018D0004). I acknowledge JPL’s High-Performance Computing (HPC) supercomputer facility, which was funded by JPL’s Information and Technology Solutions Directorate. I also acknowledge Michael Malaska, Steven Vance, and Kathryn Stack Morgan  for their suggestions during the initial data processing. I acknowledge the Galileo spacecraft mission team and the Planetary Data System (PDS) Imaging Node Server team for collecting, storing, and disseminating the SSI image and NIMS spectral data. Copyright © 2025. California Institute of Technology. Government sponsorship acknowledged.
\end{acknowledgments}

\begin{contribution}
All authors contributed equally.
\end{contribution}

\appendix

\section{Noise-removed image}
\begin{figure*}[ht!]
\plotone{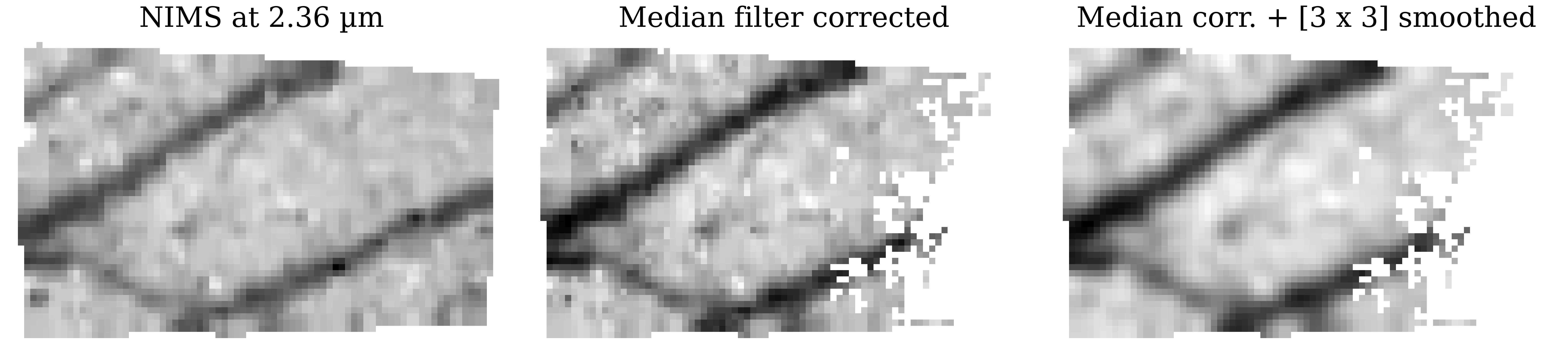}
\caption{Two-step noise removal process applied in this study\textit{.} \textit{Left panel}: Minnaert-corrected reflectance image at 2.36$\mu$m in the NIMS observation 11ENCYCLOD01A \citep{malaska2024updated}. \textit{Middle panel}: Spectral denoising using a median filter (3-band window) applied along the spectral axis at each pixel location (random noise reduction). \textit{Right panel}: Spatial smoothing applied to each spectral slice using a 3×3-pixel filter (average) on the median-filtered image (radiation noise reduction). Missing pixels in the middle and right panels result from invalid or missing data values at one or more wavelengths at the corresponding pixel locations. 
\label{fig:fig8}}
\end{figure*}

The noise-corrected image at 2.36$\mu$m in NIMS observation 11ENCYCLOD01A (Fig. \ref{fig:fig8}). 

\section{Noise-removed spectra}
\begin{figure*}[ht!]
\plotone{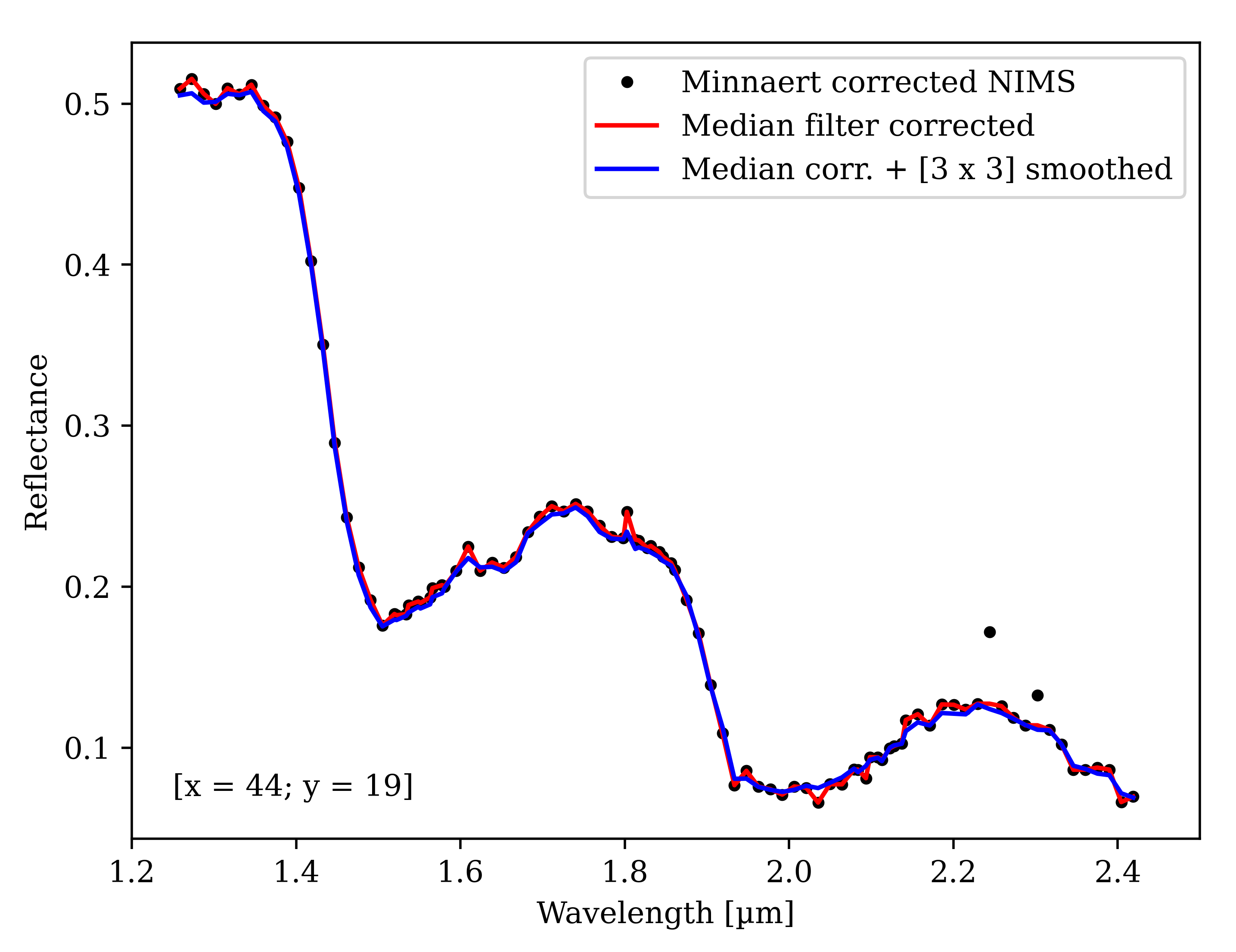}
\caption{An example spectrum following the two-step noise removal process used in this study. Black dots show the Minnaert-corrected reflectance spectrum at a single pixel \citep{malaska2018europa, malaska2024updated}. The red line represents the result after applying a 3-band median filter along the spectral axis (random noise reduced; despiked). The blue line represents the final spectrum after using a 3×3-pixel spatial smoothing filter (average) on each spectral slice of the median-filtered image (radiation noise reduced). The x, y coordinates are the location of the pixel in NIMS observation 11ENCYCLOD01A of \cite{malaska2024updated}. 
\label{fig:fig9}}
\end{figure*}

The noise-corrected spectrum at the pixel location (x: 44, y: 19) in the NIMS observation (Fig. \ref{fig:fig9}).

\section{Average reflectance of each pixel cluster}
\begin{figure*}[ht!]
\plotone{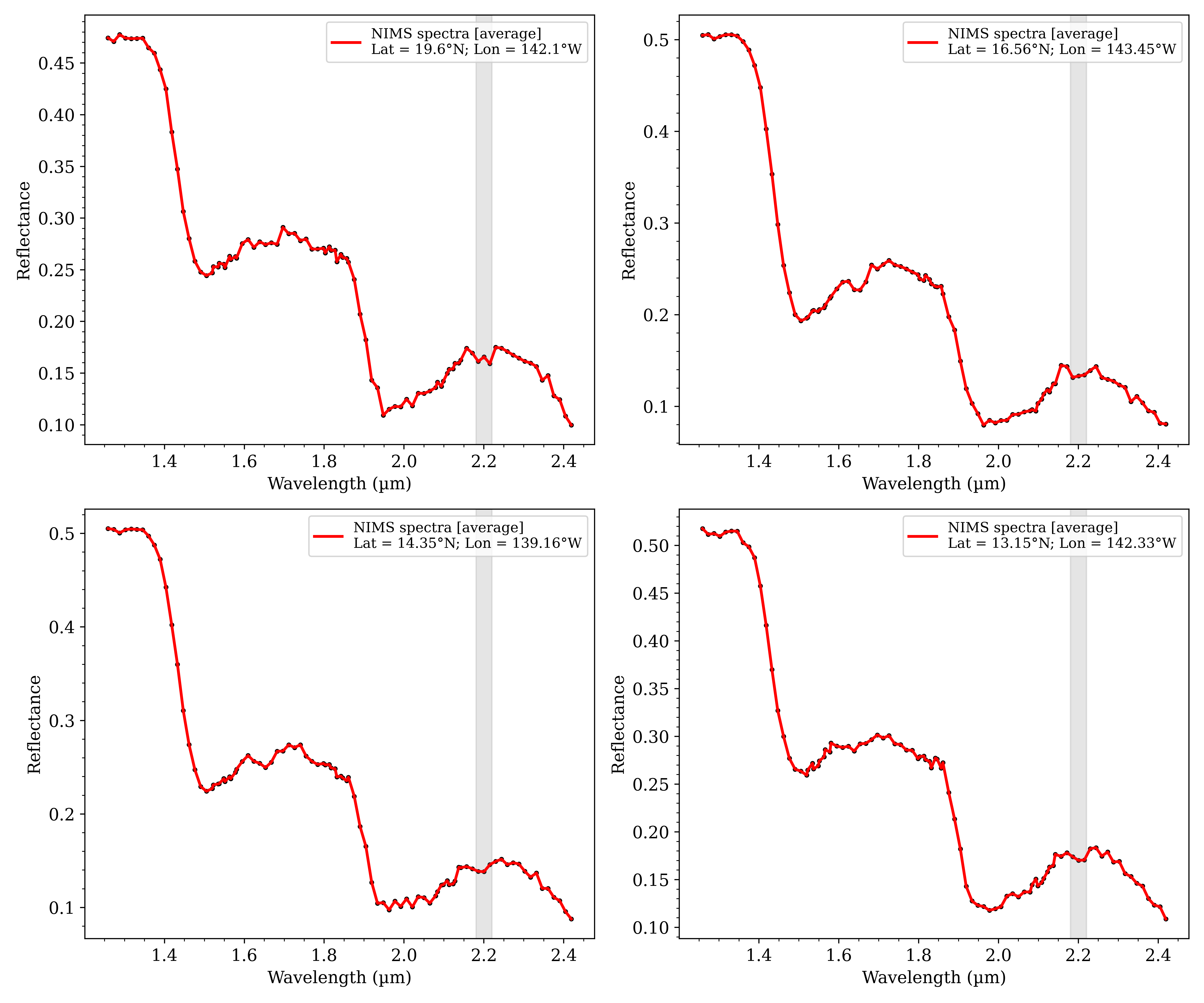}
\caption{Average reflectance spectra of each pixel cluster identified in Fig. \ref{fig:fig4}. The central latitude and longitude of each cluster are labeled in the corresponding subplot. The gray shaded region indicates the 2.18 - 2.22$\mu$m (2.20 ± 0.02$\mu$m) wavelength range, where NH\textsubscript{3}-bearing species exhibit characteristic absorption features. 
\label{fig:fig10}}
\end{figure*}

The average reflectance spectrum (Fig. \ref{fig:fig10}) of each cluster of pixels in Fig. \ref{fig:fig4}.

\section{Reflectance spectra of endmembers}
\begin{figure*}[ht!]
\plotone{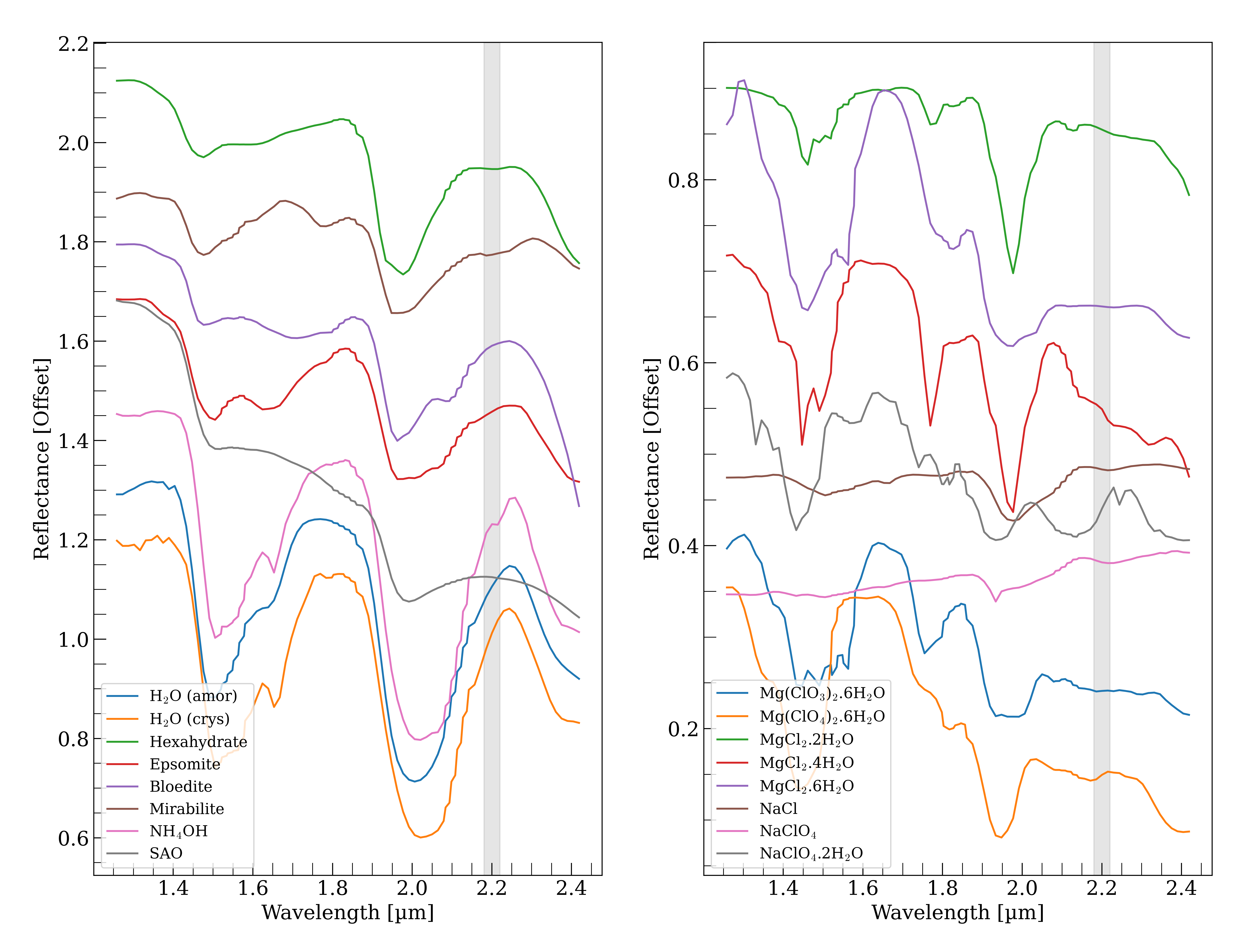}
\caption{\textit{Left panel: }The reflectance spectra (offset for clarity) of the water ice, sulfate salts, ammonia-hydrate, and SAO at near-infrared wavelengths. The modeled reflectance spectra using the \cite{shkuratov1999model} approach for a 10$\mu$m grain for amorphous and crystalline H\textsubscript{2}O ice at 120K \citep{mastrapa2008optical, mastrapa2009optical}, hexahydrite, epsomite, and bloedite at 120K \citep{dalton&pitman2012}, sulfuric acid octahydrate (SAO) at 77K \citep{carlson2005distribution}, and NH\textsubscript{3}·H\textsubscript{2}O ice at 77K \citep{brown1988search}. Mirabilite spectrum at 10$\mu$m grain was produced from the estimated optical constant at 100K \citep[][ see the main text for details]{de2021temperature}. \textit{Right panel: }The reflectance spectra from \cite{hanley2014reflectance} for magnesium chlorate (Mg(ClO\textsubscript{3})\textsubscript{2}·6H\textsubscript{2}O), magnesium perchlorate (Mg(ClO\textsubscript{4})\textsubscript{2}·6H\textsubscript{2}O), magnesium chloride (MgCl\textsubscript{2}·2H\textsubscript{2}O; MgCl\textsubscript{2}·4H\textsubscript{2}O; and MgCl\textsubscript{2}·6H\textsubscript{2}O), sodium chloride (NaCl), and sodium perchlorate (NaClO\textsubscript{4} and NaClO\textsubscript{4}·2H\textsubscript{2}O) at 80K. The gray shades of the vertical area represent the wavelengths between 2.18 and 2.22$\mu$m (2.20 ± 0.02$\mu$m), where NIMS data of the detected pixels show an absorption feature.
\label{fig:fig11}}
\end{figure*}

The reflectance spectrum of each laboratory endmember (Fig. \ref{fig:fig11}) used in the linear model.

\section{Reflectance spectra of phyllosilicate minerals}
\begin{figure*}[ht!]
\plotone{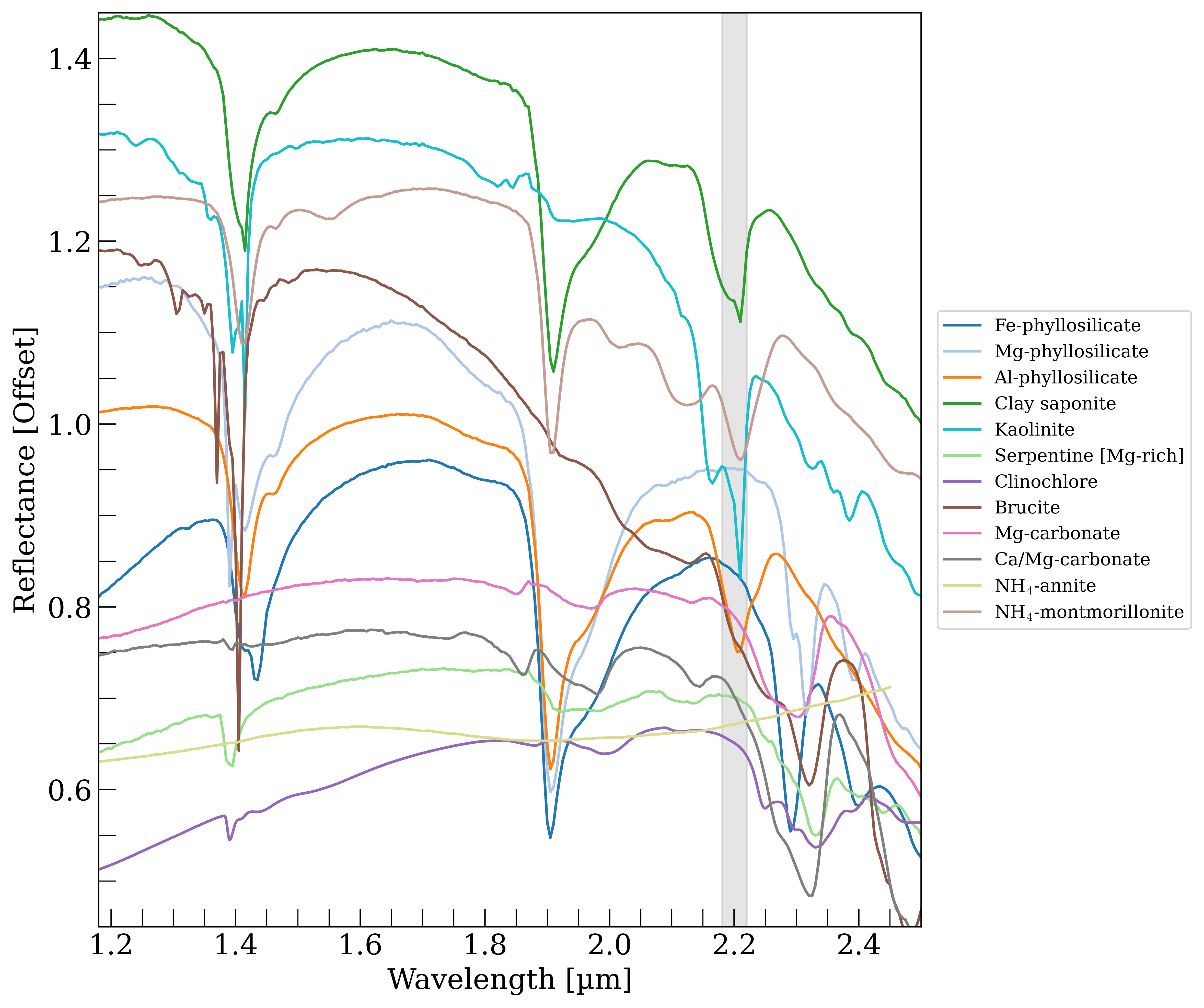}
\caption{Offset reflectance spectra of various phyllosilicate (clay) minerals in the near-infrared ($\sim$1.2–2.5$\mu$m) range, including Fe/Mg/Al-phyllosilicates, saponite, kaolinite, Mg-serpentine (lizardite), clinochlore, brucite, Mg/Ca-carbonates, and ammonium-bearing clays (NH\textsubscript{4}-annite and NH\textsubscript{4}-montmorillonite). All spectra were obtained from the RELAB spectral library \citep{milliken2021nasa}. The shaded vertical region highlights the 2.18–2.22$\mu$m wavelength range (2.20 ± 0.02$\mu$m), where the NIMS spectra of the detected pixels show an absorption feature.
\label{fig:fig12}}
\end{figure*}

 The reflectance spectrum of phyllosilicate minerals (Fig. \ref{fig:fig12}).

\section{Average and standard error data}
 The average reflectance and associated standard error data from all detected pixels (Table \ref{tab:table2}).

\begin{deluxetable*}{ccccccccc}
\tablewidth{0pt}
\tablecaption{The average reflectance and associated standard error data from all detected pixels  \label{tab:table2}}
\tablehead{
\colhead{Wavelength} & \colhead{Ref [Avg]} & \colhead{Std Err} & \colhead{Wavelength} & \colhead{Ref [Avg]} & \colhead{Std Err} & \colhead{Wavelength} & \colhead{Ref [Avg]} & \colhead{Std Err}}
\startdata
1.25853 & 0.50207 & 0.00348 & 1.65306 & 0.26258 & 0.00651 & 2.04994 & 0.11662 & 0.0044 \\
1.27304 & 0.4994 & 0.00344 & 1.66758 & 0.2667 & 0.00564 & 2.06449 & 0.1185 & 0.00488 \\
1.28755 & 0.49923 & 0.00281 & 1.68211 & 0.27469 & 0.00465 & 2.07904 & 0.12093 & 0.00461 \\
1.30205 & 0.49886 & 0.00296 & 1.69664 & 0.27885 & 0.00528 & 2.08387 & 0.12566 & 0.00498 \\
1.31656 & 0.50084 & 0.00335 & 1.71117 & 0.27904 & 0.00436 & 2.09358 & 0.12761 & 0.00533 \\
1.33107 & 0.50116 & 0.00337 & 1.72569 & 0.28055 & 0.0044 & 2.09844 & 0.12872 & 0.00449 \\
1.34558 & 0.50071 & 0.00328 & 1.74022 & 0.2754 & 0.00386 & 2.10813 & 0.13366 & 0.00472 \\
1.36008 & 0.49197 & 0.00325 & 1.75475 & 0.27288 & 0.0042 & 2.11301 & 0.13665 & 0.00456 \\
1.37459 & 0.48509 & 0.00315 & 1.76928 & 0.26705 & 0.00402 & 2.12268 & 0.14038 & 0.00451 \\
1.38909 & 0.47047 & 0.00345 & 1.7838 & 0.26545 & 0.00431 & 2.12758 & 0.14308 & 0.00504 \\
1.4036 & 0.44483 & 0.00261 & 1.79833 & 0.26228 & 0.0038 & 2.13722 & 0.14869 & 0.00423 \\
1.41811 & 0.40255 & 0.00262 & 1.80261 & 0.26039 & 0.00418 & 2.14215 & 0.15308 & 0.00511 \\
1.43261 & 0.35847 & 0.00211 & 1.81286 & 0.26152 & 0.00447 & 2.15672 & 0.16065 & 0.00406 \\
1.44712 & 0.31161 & 0.00317 & 1.81716 & 0.26037 & 0.00389 & 2.17129 & 0.16004 & 0.00425 \\
1.46162 & 0.27826 & 0.0044 & 1.82738 & 0.25847 & 0.00411 & 2.18586 & 0.1531 & 0.00473 \\
1.47613 & 0.25301 & 0.00521 & 1.83171 & 0.25088 & 0.00383 & 2.20043 & 0.15331 & 0.00426 \\
1.49063 & 0.23718 & 0.00634 & 1.84191 & 0.25517 & 0.00471 & 2.215 & 0.15355 & 0.00397 \\
1.50514 & 0.23294 & 0.00677 & 1.84626 & 0.25367 & 0.00484 & 2.22957 & 0.16291 & 0.00475 \\
1.51965 & 0.2335 & 0.00627 & 1.85643 & 0.25006 & 0.00408 & 2.24413 & 0.16469 & 0.00464 \\
1.5223 & 0.23755 & 0.00661 & 1.86081 & 0.24947 & 0.00497 & 2.2587 & 0.1568 & 0.00467 \\
1.53415 & 0.24178 & 0.00656 & 1.87536 & 0.2252 & 0.00472 & 2.27327 & 0.15725 & 0.00502 \\
1.53683 & 0.24098 & 0.00629 & 1.88991 & 0.19889 & 0.0037 & 2.28784 & 0.15243 & 0.00442 \\
1.54866 & 0.24245 & 0.00662 & 1.90446 & 0.1702 & 0.00373 & 2.30241 & 0.14938 & 0.00491 \\
1.55136 & 0.24335 & 0.00663 & 1.91901 & 0.1337 & 0.00268 & 2.31697 & 0.14306 & 0.00449 \\
1.56316 & 0.24868 & 0.00696 & 1.93356 & 0.11865 & 0.00343 & 2.33154 & 0.1379 & 0.00532 \\
1.56589 & 0.25074 & 0.00704 & 1.94811 & 0.10825 & 0.00325 & 2.34611 & 0.13125 & 0.00431 \\
1.57766 & 0.25396 & 0.00627 & 1.96265 & 0.10438 & 0.00417 & 2.36068 & 0.12935 & 0.00454 \\
1.58041 & 0.25764 & 0.00684 & 1.9772 & 0.10688 & 0.00375 & 2.37524 & 0.11671 & 0.00402 \\
1.59494 & 0.26367 & 0.00616 & 1.99175 & 0.10544 & 0.00414 & 2.38981 & 0.11256 & 0.00367 \\
1.60947 & 0.26726 & 0.00536 & 2.0063 & 0.11003 & 0.00405 & 2.40437 & 0.10296 & 0.00391 \\
1.624 & 0.26508 & 0.00548 & 2.02085 & 0.11053 & 0.00487 & 2.41894 & 0.09527 & 0.00307 \\
1.63853 & 0.2618 & 0.00578 & 2.03539 & 0.11792 & 0.00463 & & &	
\enddata
\end{deluxetable*}

\bibliography{Emran_2025}{}
\bibliographystyle{aasjournalv7}
\end{document}